\documentclass[12pt]{article}

\global\arraycolsep=1pt

\usepackage{color,amsbsy,amssymb,latexsym,amsfonts, amsmath,cancel}
\usepackage{braket}
\usepackage{mathrsfs}
\usepackage{graphicx}
\usepackage{cite}

\newcommand{\bel}[1]{\begin{equation}\label{#1}}                     
\newcommand{\bal}[1]{\begin{eqnarray}\label{#1}}   
\newcommand{\be}{\begin{equation}}               
\newcommand{\ba}{\begin{eqnarray}}           
\newcommand{\ee}{\end{equation}}
\newcommand{\ea}{\end{eqnarray}}

\renewcommand{\thefootnote}{\fnsymbol{footnote}}

\newcommand{\bea}{\begin{equation}}
\newcommand{\eea}{\end{equation}}
\newcommand\fD{\mathfrak D}

\begin{document}

%
%
\begin{titlepage}
\begin{flushright}
OCU-PHYS 441\\
\end{flushright}

\vspace{15pt}

\begin{center}
{\Large\bf 
Nambu-Jona-Lasinio Theory and \\ Dynamical Breaking of Supersymmetry}
\end{center}

\vspace{13pt}

\begin{center}
{\large Nobuhito Maru }\\
%
\vspace{18pt}
%

 \it Department of Mathematics and Physics, 
Osaka City University, Osaka 558-8585, Japan

\end{center}
%
\vspace{20pt}
\begin{center}
Abstract\\
\end{center}
%
A recently proposed new mechanism of D-term triggered dynamical supersymmetry breaking is reviewed. 
Supersymmetry is dynamically broken by nonvanishing D-term vacuum expectation value, 
which is realized as a nontrivial solution of the gap equation in the self-consistent approximation 
as in the case of Nambu-Jona-Lasinio model and BCS superconductivity.

\vfill

\setcounter{footnote}{0}
\renewcommand{\thefootnote}{\arabic{footnote}}

\end{titlepage}

\renewcommand{\thefootnote}{\arabic{footnote}}
\setcounter{footnote}{0}

\section{Introduction}
Supersymmetry (SUSY) is one of the attractive solutions to the hierarchy problem, 
but it has to be broken spontaneously at low energy because of undiscovered superparticles. 
SUSY should be broken nonperturbatively or dynamically \cite{Witten1, Witten2} according to nonrenormalization theorem \cite{GRS}. 
Although the mechanism of dynamical SUSY breaking (DSB) by F-term has been much explored \cite{ADS1,ADS2,ADS3,Veneetal1,Veneetal2}, 
models of DSB by D-term were not known. 
Several years ago, such a simple mechanism of D-term DSB (DDSB) was proposed 
by H. Itoyama and the present author \cite{imaru1, imaru2, imaru3} 
in which the nonvanishing vacuum expectation value (VEV) of D-term is dynamically realized 
as a nontrivial solution of the gap equation in the self-consistent Hartree-Fock approximation 
as in the case of Nambu-Jona-Lasinio (NJL) model \cite{NJL1, NJL2} and BCS superconductivity \cite{BCS1, BCS2}. 
In our mechanism, the gauge sector is extended to be ${\cal N} = 2$ supersymmetric, 
and gaugino becomes massive by the D-term VEV through the Dirac mass term with ${\cal N}=2$ partner fermion of gaugino. 
Thus, our mechanism can be directly applied to Dirac gaugino scenario \cite{FNW} 
which is an interesting alternative extension of the minimal SUSY Standard Model (MSSM).  
Much attentions have been paid to various phenomenological studies based on Dirac gaugino scenario and its extensions 
\cite{Fayet, PS, HR, NRSU, KPW, Benakli1, Benakli2, Benakli3, Benakli4, Benakli5, Benakli6, Benakli7, Carpenter, CCFKZ, AG, ABFP, FK, DMRM, KP, KM, YMFJK, FKMT, 
FGKP, Unwin, CJPS, DGHT, GT, Busbridge, NR, DLSTZ, AGWM, Martin, GKMPS, CDHRW, DSZ, MNS, NY, CCG}. 

This paper is organized as follows. 
In the next section, we start out from exhibiting the component action from that of the superspace, 
state the set of assumptions we have made.  
We review the original reasoning that has led us to the D-term triggered DSB. 
We set up the background field formalism to be used in the subsequent sections, 
separating the three kinds of background from the fluctuations. 
In section three, we elaborate upon our treatment of the effective potential with the three kinds of background fields 
as well as the point of the Hartree-Fock approximation in Refs. \cite{imaru1, imaru2,imaru3}. 
Section four is the main thrust of this paper. 
We present our variational analyses of the effective potential in full detail. 
Treating one of the order parameters F-term as an induced perturbation, 
we demonstrate that the stationary values of D-term and an adjoint scalar field are determined by the intersection of the two real curves, 
namely, the simultaneous solution to the gap equation and the equation of stationarity for the adjoint scalar field. 
Numerical analysis is provided that demonstrates the existence of such solution as well as the self-consistency of our analysis. 
The second variation of the scalar potential is computed and the local stability of the vacuum is shown from the numerical data. 
Section five will discuss the lifetime of our SUSY breaking vacuum. 
Unlike non-SUSY case, our SUSY breaking vacuum is necessarily meta-stable 
because of positive semi-definiteness of the vacuum energy in the rigid SUSY theory. 
Namely, the vacuum energy of our SUSY breaking vacuum is higher than the SUSY vacuum 
which is a trivial solution of the gap equation. 
We will show that the lifetime of our vacuum can be sufficiently large by adjusting parameters in the theory. 
In section six, a realization of observed Higgs mass by extra $U(1)$ D-term contributions to Higgs mass will be discussed \cite{imaru4}. 
Summary is given in the last section.

\section{The action, assumptions and some properties}
The action we discuss is the general ${\cal N}=1$ supersymmetric action 
consisting of chiral superfield $\Phi^a=(\phi^a, \psi^a, F^a)$ in the adjoint representation 
and the vector superfield $V^a=(\lambda^a, V_\mu^a, D^a)$ with three input functions, 
the K\"{a}hler potential $K(\Phi^a, \bar{\Phi}^a)$ with its gauging, the gauge kinetic superfield $\tau_{ab}(\Phi^a)$
 that follow from the second derivatives of a generic holomorphic function ${\cal F}(\Phi^a)$,
and the superpotential $W(\Phi^a)$, 
\ba
    {\cal L}
     =     
           \int d^4 \theta K(\Phi^a, \bar{\Phi}^a) + (gauging)   
        + \int d^2 \theta
           {\rm Im} \frac{1}{2} 
           \tau_{ab}(\Phi^a)
           {\cal W}^{\alpha a} {\cal W}^b_{\alpha}
            + \left(\int d^2 \theta W(\Phi^a)
         + c.c. \right).       
           \label{KtauW}
\ea
The gauge group is taken to be $U(N)$ and, for simplicity, 
 we assume that the theory is in the unbroken phase of the entire gauge group, 
 which can be accomplished by tuning the superpotential. 
We also assume that third derivatives of ${\cal F}(\Phi^a)$ at the scalar vacuum expectation values (VEV's) are non-vanishing. 


The component Lagrangian of Eq. (\ref{KtauW}) reads 
\ba
{\cal L}_{U(N)} = {\cal L}_{{\rm K\ddot{a}hler}} + {\cal L}_{{\rm gauge}} + {\cal L}_{{\rm sup}},
\ea
where
\ba
 {\cal L}_{{\rm K\ddot{a}hler}}
    &=&    g_{ab} {\cal D}_\mu \phi^a {\cal D}^\mu \bar{\phi}^b
         - \frac{i}{2} g_{ab} \psi^a \sigma^\mu {\cal D}_\mu' \bar{\psi}^b
         + \frac{i}{2} g_{ab} {\cal D}_\mu' \psi^a \sigma^\mu \bar{\psi}^b + g_{ab} F^a \bar{F}^b
           \nonumber \\
    & &  - \frac{1}{2} g_{ab,\bar{c}} F^a \bar{\psi}^b \bar{\psi}^c
         - \frac{1}{2} g_{bc,a} \bar{F}^c \psi^a \psi^b
         + \frac{1}{\sqrt{2}} g_{ab} (\lambda^c \psi^a k_c^*{}^{b} 
         + \bar{\lambda}^c \bar{\psi}^b k_c{}^{a})
         + \frac{1}{2} D^a \fD_a,
           \label{L:K} \\
  {\cal L}_{{\rm gauge}}
    &=&  - \frac{1}{2} {\cal F}_{ab} \lambda^a \sigma^\mu {\cal D}_\mu \bar{\lambda}^b
         - \frac{1}{2} \bar{{\cal F}}_{ab} {\cal D}_\mu \lambda^a \sigma^\mu \bar{\lambda}^b
         - \frac{1}{4} 
         g_{ab} F_{\mu \nu}^a F^{b \mu \nu}
         - \frac{1}{8} ({\rm Re} {\cal F})_{ab} \epsilon^{\mu \nu \rho \sigma} 
           F_{\mu \nu}^a F_{\rho \sigma}^b
           \nonumber \\
    & &  - \frac{\sqrt{2} i}{8} ({\cal F}_{abc} \psi^c \sigma^\nu \bar{\sigma}^\mu \lambda^a
         - \bar{{\cal F}}_{abc} \bar{\lambda}^a \bar{\sigma}^\mu \sigma^\nu \bar{\psi}^c ) F_{\mu \nu}^b
                 + \frac{\sqrt{2}}{4} ( {\cal F}_{abc} \psi^c\lambda^a
         + \bar{{\cal F}}_{abc} \bar{\psi}^c \bar{\lambda}^a ) D^b
                    \nonumber \\
    & &  
+ \frac{1}{2}   
g_{ab}  D^a D^b 
         + \frac{i}{4} {\cal F}_{abc} F^c \lambda^a \lambda^b
         - \frac{i}{4} \bar{{\cal F}}_{abc} \bar{F}^c \bar{\lambda}^a \bar{\lambda}^b
           \nonumber \\
    & &  
- \frac{i}{8} {\cal F}_{abcd} \psi^c \psi^d \lambda^a \lambda^b
         + \frac{i}{8} \bar{{\cal F}}_{abcd} \bar{\psi}^c \bar{\psi}^d \bar{\lambda}^a \bar{\lambda}^b,
           \label{L:gauge} \\
     {\cal L}_{{\rm sup}}
    &=&    F^a \partial_a W
         - \frac{1}{2} \partial_a \partial_b W \psi^a \psi^b
         + c.c.,
           \label{L:superpotential}        
\ea
where
\ba
\fD_a = -\frac{1}{2} ({\cal F}_b f_{ac}^b \bar{\phi}^c + \bar{\cal F}_b f^b_{ac} \phi^c)
\ea
and $f_{ac}^b$ is the structure constant of $SU(N)$. 
Note that an equation of motion for $F^a$ is $F^a = -g^{ab} \overline{\partial_bW}$ + fermions. 
We also assume $\langle F^a \rangle_{{\rm tree}} = -\langle g^{ab} \overline{\partial_b W} \rangle_{{\rm tree}} =0$ at the tree level.  
At the lowest order in perturbation theory, 
there is no source which gives VEV to the auxiliary field $D^0$: $\langle D^0 \rangle_{{\rm tree}} = 0$. 
The $U(N)$ gaugino is massless at the tree level while the fermionic partner of the scalar gluon receives 
 the tree level mass $m_a = m_0 = \langle g^{00} \partial_0 \partial_0 W \rangle_{{\rm tree}} $.

\subsection{Original reasoning of DDSB}

In Ref. \cite{imaru1}, it was shown that the VEV of an auxiliary field $D^0$ is non-vanishing in the Hartree-Fock approximation. 
Therefore, the theory realizes the D-term dynamical supersymmetry breaking. 

The part of the Lagrangian providing the fermion mass matrix of size $2N$ 
is
\ba
-\frac{1}{2} (\lambda^a, \psi^a) 
\left(
\begin{array}{cc}
0 & -\frac{\sqrt{2}}{4} {\cal F}_{abc} D^b \\
-\frac{\sqrt{2}}{4} {\cal F}_{abc} D^b & \partial_a \partial_c W \\
\end{array}
\right) 
\left(
\begin{array}{c}
\lambda^c \\
\psi^c \\
\end{array}
\right) + (c.c.).
\label{massterm}
\ea 
 
 It was observed that the auxiliary $D^a$ field,
  which is an order parameter of ${\cal N} =1$ supersymmetry,
    couples to the fermionic (but not bosonic) bilinears through the third prepotential derivatives:
     the non-vanishing VEV of $D^0$ immediately gives a Dirac mass of the fermions.
The equation of motion for the auxiliary field $D^0$ implies 
  \ba
    \langle D^{0} \rangle
    =    - \frac{1}{2 \sqrt{2}} \langle g^{00} 
           \left( {\cal F}_{0cd}\psi^d \lambda^c 
         + \bar{{\cal F}}_{0cd} \bar{\psi}^d \bar{\lambda}^c \right) \rangle,
 \ea
 telling us that the condensation of the Dirac bilinear is responsible for $\langle D^{0} \rangle \neq 0$. 
This feature reminds us of the electromagnetic $U(1)$ symmetry breaking in BCS theory by Cooper pair condensation
 or the chiral symmetry breaking in QCD by the quark-antiquark condensation. 
 
 We diagonalize the holomorphic part of the mass matrix: 
\ba
  M_{F a} \equiv
\left(
\begin{array}{cc}
0 & -\frac{\sqrt{2}}{4} \langle {\cal F}_{0aa} D^0 \rangle \\
-\frac{\sqrt{2}}{4} \langle {\cal F}_{0aa} D^0 \rangle & \langle \partial_a \partial_a W \rangle \\
\end{array}
\right). 
\label{massmatrix1}
\ea 
 Note that the non-vanishing third prepotential derivatives are ${\cal F}_{0aa}$
 where $a$ refers to the generators of the unbroken gauge group.
 By an orthogonal transformation, we obtain the two eigenvalues of Eq. (\ref{massmatrix1}) for each generator, 
 which are mixed Majorana-Dirac type  :
 \ba
 \label{eigenvalue}
 {\Lambda}_{a {\bf 11}}^{(\pm)} &=&  \frac{1}{2} \langle \partial_a \partial_a W \rangle 
 \left( 1 \pm \sqrt{1 + \frac{\langle {\cal F}_{0aa}D^0 \rangle^2}{2\langle \partial_a \partial_a W\rangle^2} }\right). 
  \ea
Introducing 
\ba
 \lambda_{a{\bf 11}}^{(\pm)} \equiv \frac{1}{2}\left( 1 \pm \sqrt{1 +\Delta_{{\bf 11}}^2} \right), \quad
  \Delta_{a{\bf 11}}^2 \equiv   \frac{\langle {\cal F}_{0aa} D^0 \rangle^2}{2 \langle \partial_a \partial_a W \rangle^2},
  \label{lambdapm}
\ea
 we obtain
\ba
 | \Lambda_{a {\bf 11}}^{(\pm)} |^2 = | \langle \partial_a \partial_a W \rangle | | \lambda_{a{\bf 11}}^{(\pm )} |^2.
\ea
 
 It was also shown in ref. \cite{imaru1} that the non-vanishing $F^0$ term is induced
 by the consistency of our procedure of computation. (See also \cite{DNNS, CFK}). 
  This is because the stationary value of the scalar fields  gets shifted upon the variation (the vacuum condition). 
  The final mass formula for the $SU(N)$ fermions is to be read off from  
\ba
{\cal L}_{{\rm mass}}^{(holo)} 
= -\frac{1}{2} \langle g_{0a,a} \rangle \langle \bar{F}^0 \rangle \psi^a \psi^a 
+ \frac{i}{4} \langle {\cal F}_{0aa} \rangle \langle F^0 \rangle \lambda^a \lambda^a
-\frac{1}{2} \langle \partial_a \partial_a W \rangle \psi^a \psi^a 
+ \frac{\sqrt{2}}{4}
\langle {\cal F}_{0aa} \rangle \psi^a \lambda^a 
\langle D^0 \rangle  \nonumber \\
\ea
  We will write down the explicit form in the next subsection. 
 See Eqs. (\ref{imaru3massmatrix}), (\ref{parameters}), (\ref{ev1}) and (\ref{ev2}).
A main remaining point is how to establish the procedure in which the stationary values of the scalar fields,
 $D^0$ and $F^0$ perturbatively induced are determined, which we will resolve in this paper.

\subsection{Quadratic part of the quantum action}
 In this subsection, we write down parts of the action with the background fields for the computation of the one-loop determinant
  in the next section. 
  
\subsubsection{Fermionic part}
Let us extract the fermion bilinears from Eqs. (\ref{L:K}), (\ref{L:gauge}) and (\ref{L:superpotential}) 
which are needed for our analysis in what follows. 
Rescaling the fermion fields so that their kinetic terms become canonical, we obtain
\ba
{\cal L}_F 
&=& - \frac{i}{2} \psi^a \sigma^\mu \partial_\mu \bar{\psi}^a
         + \frac{i}{2} (\partial_\mu \psi^a) \sigma^\mu \bar{\psi}^a 
 - \frac{i}{2} \lambda^a \sigma^\mu \partial_\mu \bar{\lambda}^a
         + \frac{i}{2} (\partial_\mu \lambda^a) \sigma^\mu \bar{\lambda}^a \nonumber \\
    & &  - \frac{1}{2} \left( g^{bb} g_{0b,\bar{b}} F^0 \right) \bar{\psi}^b \bar{\psi}^b
         - \frac{1}{2} \left( g^{bb} g_{0b,b} \bar{F}^0 \right) \psi^b \psi^b \nonumber \\
    &&     + \frac{\sqrt{2}}{4} \left( {\cal F}_{0aa} \sqrt{g^{aa}~{\rm Im}{\cal F}^{aa}} D^0 \right) \psi^a \lambda^a
         + \frac{\sqrt{2}}{4} \left( \bar{{\cal F}}_{0aa} \sqrt{g^{aa}~{\rm Im}{\cal F}^{aa}} D^0 \right) \bar{\psi}^a \bar{\lambda}^a 
         \nonumber \\
        && + \frac{i}{4} \left( {\cal F}_{0aa} g^{aa} F^0 \right) \lambda^a \lambda^a
         - \frac{i}{4} \left( \bar{{\cal F}}_{0aa} g^{aa} \bar{F}^0 \right) \bar{\lambda}^a \bar{\lambda}^a \nonumber \\
      &&   - \frac{1}{2} \left( g^{aa} \partial_a \partial_a W \right) \psi^a \psi^a 
         -\frac{1}{2} \left( g^{aa} \overline{\partial_a \partial_a W} \right) \bar{\psi}^a \bar{\psi}^a. 
\ea
Here the fermion fields $\psi^a$, $\bar{\psi}^a$, $\lambda^a$, $\bar{\lambda}^a$ are to be integrated to make a part of the effective potential, 
while the gauge kinetic function ${\cal F}_{aa}$, the K\"ahler metric $g_{aa}$ and their derivatives are functions of 
the $U(N)$ singlet $c$-number background scalar field $\varphi^0$. 
The order parameters of supersymmetry $F^0$, $\bar{F}^0$, and $D^0$ are taken as background fields as well. 

From the lagrangian ${\cal L}_F$, the holomorphic part of the mass matrix is read off as
\ba
{\cal M}_a =
\left(
\begin{array}{cc}
-\frac{i}{2} g^{aa} {\cal F}_{0aa} F^0 & -\frac{\sqrt{2}}{4} \sqrt{g^{aa} ({\rm Im}{\cal F})^{aa}}{\cal F}_{0aa} D^0 \\
-\frac{\sqrt{2}}{4} \sqrt{g^{aa} ({\rm Im}{\cal F})^{aa}}{\cal F}_{0aa} D^0 & g^{aa} \partial_a \partial_a W + g^{aa} g_{0a,a} \bar{F}^0 \\
\end{array}
\right) 
= \left(
\begin{array}{cc}
m_{\lambda\lambda}^a & m_{\lambda\psi}^a \\
m_{\psi \lambda}^a  
& m_{\psi\psi}^a \\
\end{array}
\right). 
\label{imaru3massmatrix}
\ea
We parametrize this matrix such that, in the case of $F^0=\bar{F}^0=0$, its form reduces to that of Ref. \cite{imaru1, imaru2}.
The quantities with multiple indices such as ${\cal F}_{0aa}$ receive $U(N)$ invariant expectation values:
$\langle {\cal F}_{0aa} \rangle = \langle {\cal F}_{000} \rangle$ e.t.c.
We suppress the indices as we work with the unbroken $U(N)$ phase in this paper.
\ba
\Delta \equiv -  \frac{2m_{\lambda\psi}}{m_{\psi\psi}}, \qquad f \equiv \frac{2im_{\lambda\lambda}}{{\rm tr}{\cal M}}. 
\label{parameters}
\ea
The two eigenvalues of the holomorphic mass matrix are written as
\ba
\Lambda^{(\pm)} \equiv ({\rm tr}{\cal M}) \lambda^{(\pm)},
\label{ev1}
\ea
where
\ba
\lambda^{(\pm)} = \frac{1}{2} \left( 1 \pm \sqrt{(1+if)^2 + \left( 1+\frac{i}{2}f \right)^2 \Delta^2 } \right).
\label{ev2}
\ea
 These provide the masses for the two species of $SU(N)$ fermions once the stationary
values are determined.

\subsubsection{Bosonic part}

Next, we extract the bosonic quantum bilinears from Eqs. (\ref{L:K}), (\ref{L:gauge}), and (\ref{L:superpotential}). 
Let 
\ba
\phi^a &=& \delta_0^a \varphi^0 + \sqrt{g^{aa}(\varphi)} \tilde{\varphi}^a, \\ 
A_\mu^a &=& \sqrt{({{\rm Im}~{\cal F}})^{aa}} \tilde{A}_\mu^a, \\
F^a &=& \sqrt{g^{aa}(\varphi)} \tilde{F}^a, \\
D^a &=& \sqrt{({\rm Im}~{\cal F})^{aa}} \tilde{D}^a
\ea
where $\varphi^0$ are the background field 
while $\tilde{\varphi}^a$, $\tilde{A}_\mu^a$, $\tilde{F}^a$ and $\tilde{D}^a$ are the quantum scalar, vector 
and auxiliary fields respectively. 

We obtain
\ba
{\cal L}_B^{(1)} &=& \partial_\mu \tilde{\varphi}^a \partial^\mu \tilde{\varphi}^{*a} 
-\frac{1}{4} \tilde{F}_{\mu \nu}^a \tilde{F}^{a \mu \nu} 
+ \tilde{F}^a \tilde{\bar{F}}^a + \frac{1}{2} \tilde{D}^a \tilde{D}^a \nonumber \\
&&+ \tilde{F}^a \left( (\sqrt{g^{aa}} \partial_a W) + (g^{aa} \partial_a \partial_a W) \tilde{\varphi}^a \right)
+ \tilde{\bar{F}}^a \left( (\sqrt{g^{aa}} \overline{\partial_a W}) + (g^{aa} \overline{\partial_a \partial_a W}) \tilde{\varphi}^{a*} \right).
\ea
We have also ignored 
$-\frac{1}{8}({\rm Re}~{\cal F})_{ab}\epsilon^{\mu\nu\rho\sigma}F_{\mu\nu}^a F_{\rho\sigma}^b$ 
as we eventually set $\varphi^a$ to be constant in our analysis and this term becomes a total derivative.

\subsection{Connection with the previous work}
We here stop shortly to address the connection of ref. \cite{imaru1} with the previous work. 
Models of dynamical supersymmetry breaking with non-vanishing F- and D-terms 
have been previously proposed: they are, for instance, the 3-2 model \cite{ADS3} 
and the 4-1 model in \cite{DNNS}. 
In these models, supersymmetry is unbroken at the tree level and 
  is broken by the non-vanishing VEV of the F-term through instanton generated superpotentials. 
Non-vanishing VEV of the D-term  is also induced, but is much smaller than that of the F-term. 

In our mechanism, supersymmetry is unbroken at the tree level, 
 and is broken in a self-consistent Hartree-Fock approximation of  the NJL type
  that produces a non-vanishing VEV for the D-term.
 A non-vanishing VEV for the F-term is induced in our Hartree-Fock vacuum that shifts the tree vacuum
  and we explore the region of the parameter space in which F-term VEV is treated perturbatively.
 
  We should mention that  the way in which the two kinds of gauginos (or the gaugino and the adjoint matter fermion)
   receive masses    is  an extension of that proposed in \cite{FNW}: 
 the pure Dirac-type gaugino mass is generated in \cite{FNW} 
while the mixed Majorana-Dirac type gaugino masse is generated in our case, the Majorana part being given by the
 second derivative of the superpotential. 
In \cite{FNW}, the dynamical origin of non-vanishing D-term VEV was not addressed.

 As for the application to dynamical chiral symmetry breaking,  
 a supersymmetric NJL type model has been considered \cite{BL, CLB, MR, FJK}. 
Chiral symmetry is not spontaneously broken in a supersymmetric case. 
Even in softly broken supersymmetric theories, 
the chiral symmetry broken phases are degenerate with the chirally symmetric ones. 
Thus, in supersymmetric theories, the phase with broken chiral symmetry is no longer the energetically preferred ground state.

\section{The effective potential in the Hartree-Fock approximation}

The goal of this section is to determine the effective potential to the leading order in the Hartree-Fock approximation. 
We will regulate one-loop integral by the dimensional reduction \cite{Siegel}. 
We prepare a supersymmetric counterterm, and  set the normalization condition. 
We make brief comments on regularization and  subtraction schemes in the end of section 4. 
 We also change the notation for  expectation values in general from $\langle ... \rangle$ to $..._*$
  as our main thrust of this paper is the determination of the stationary values from the variational analysis.


In the Hartree-Fock approximation, one begins with considering the situation 
where one-loop corrections in the original expansion in $\hbar$ become large 
and are comparable to the tree level contribution. 
The optimal configuration of the effective potential to this order is found 
by matching the tree against one-loop, varying with respect to the auxiliary fields. 
We start the analysis of this kind for our effective potential. 
There are three constant background fields as arguments of the effective potential: 
$\varphi \equiv \varphi^0~({\rm complex})$, $U(N)$ invariant background scalar, 
$D \equiv D^0~({\rm real})$ and $F \equiv F^0~({\rm complex})$. 
The latter two are the order parameters of ${\cal N}=1$ supersymmetry. 

We vary our effective potential with respect to all these constant fields and examine the stationary conditions. 
We  also examine a second derivative at the stationary point along the constraints of the auxiliary fields 
to understand better  the Hartree-Fock corrected mass of the scalar gluons. 
Let us denote our effective potential by $V$. 
It consists of three parts: 
\ba
V = V^{{\rm tree}} + V_{{\rm c.t.}} + V_{{\rm 1-loop}}. 
\ea
The first term is the tree contributions, the second one is the counterterm and the last one is the one-loop contributions. 
After the elimination of the auxiliary fields, 
the effective potential is referred to as the scalar potential so as to be distinguished from the original $V$. 

\subsection{The tree part}

To begin with, let us write down the tree part and find a parametrization by two complex and one real parameters. 
We also introduce simplifying notation $g_{00}(\varphi, \bar{\varphi}) \equiv g(\varphi, \bar{\varphi}), 
({\rm Im}~{\cal F}(\varphi))_{00} \equiv {\rm Im}~{\cal F}''(\varphi), \partial_0 W(\varphi) = W'(\varphi), g_{00,0} \equiv \partial g,$ etc. 
\ba
V^{{\rm tree}}(D, F, \bar{F}, \varphi, \bar{\varphi}) =-gF\bar{F} -\frac{1}{2} ({\rm Im}{\cal F}'') D^2 - F W' - \bar{F}\bar{W'}.  
\label{treepart}
\ea
As a warm up, let us determine the vacuum configuration by a set of stationary conditions at the tree level:
\ba
&&\frac{\partial V^{{\rm tree}}}{\partial D} = 0, \label{dflattree} \\
&&\frac{\partial V^{{\rm tree}}}{\partial F} = \frac{\partial V^{{\rm tree}}}{\partial \bar{F}} = 0, \label{fflattree} \\
&&\frac{\partial V^{{\rm tree}}}{\partial \varphi} = \frac{\partial V^{{\rm tree}}}{\partial \bar{\varphi}} = 0.  \label{phistattree}
\ea
Eq. (\ref{dflattree}) determines the stationary value of $D$:
\ba
D=0 \equiv D_*, 
\ea
while from Eq. (\ref{fflattree}), we obtain 
\ba
F=-g^{-1}(\varphi, \bar{\varphi}) \bar{W}'(\bar{\varphi}) \equiv F_*(\varphi, \bar{\varphi}).
\ea
Eq. (\ref{phistattree}) together with these two gives
\ba
W'(\varphi_*) = 0,~{\rm i.e.}~F_*(\varphi, \bar{\varphi})=0,
\ea
as well as
\ba
V_{{\rm scalar}}^{{\rm tree}} (\varphi, \bar{\varphi}) \equiv V^{{\rm  tree}} (\varphi, \bar{\varphi}, D_*=0, F=F_*(\varphi, \bar{\varphi}), 
\bar{F}=\overline{F_*(\varphi, \bar{\varphi})}) = g^{-1}(\varphi, \bar{\varphi}) |W'(\varphi)|^2. 
\ea
The negative coefficients of the RHS of Eq. (\ref{treepart}) imply that both $D$ and $F$ profiles of the potential
have a maximum for a given $\varphi$. These signs are, of course, the right signs for the stability of the scalar potential
 as is clear by completing the square. This is a trivial comment to make here but will become less trivial later.
The mass of the scalar gluons at tree level $|m_{s*}|^2$ is read off from the second derivative at the stationary point:
\ba
&&\left. 
\frac{\partial^2 V^{{\rm tree}}(\varphi, \bar{\varphi})}{\partial \varphi \partial \bar{\varphi}}
\right|_{\varphi_*, \bar{\varphi}_*} 
= g^{-1}(\varphi_*, \bar{\varphi}_*) \left| W''(\varphi_*)\right|^2, \\
&&m_s(\varphi, \bar{\varphi}) \equiv g^{-1}(\varphi, \bar{\varphi}) W''(\varphi), \label{sgluonmass} \quad
m_{s*} = m_s (\varphi_*, \bar{\varphi}_*).  
\ea
As we have already introduced in Eq. (\ref{parameters}), $\Delta$ and $r$ are defined by 
\ba
\Delta \equiv -2 \frac{m_{\lambda \psi}}{m_{\psi \psi}} = \frac{\sqrt{2}}{2} 
\frac{\sqrt{g^{-1}({\rm Im}{\cal F}'')^{-1}} {\cal F}'''}{g^{-1}W''+g^{-1}\partial g \bar{F}} D
\equiv r(\varphi, \bar{\varphi}, F, \bar{F}) D.
\ea
Recall that we have suppressed the indices, invoking the $U(N)$ invariance of the expectation values. 
Also
\ba
f_3 \equiv \frac{g^{-1}{\cal F}''' F}{g^{-1}W'' + g^{-1} \partial g \bar{F}}, 
\label{f3}
\ea
where $f_3$ differs from $f$ in Eq. (\ref{parameters}) by
\ba
(g^{-1}W'' + g^{-1} \partial g \bar{F}) f_3 = \left( g^{-1}W'' + g^{-1} \partial g \bar{F} -\frac{i}{2}g^{-1}{\cal F}'''F \right)f. 
\ea 
We obtain 
\ba
F=\frac{m_s}{g^{-1}{\cal F}'''} \varepsilon, \qquad \bar{F}=\frac{\bar{m}_s}{g^{-1} \bar{{\cal F}}'''} \varepsilon, \qquad 
\varepsilon = \frac{f_3 + \frac{\bar{m}_s}{m_s} \frac{g^{-1}\partial g}{g^{-1}\bar{{\cal F}}'''}  |f_3|^2}{1 - \left| \frac{g^{-1}\partial g f_3}{g^{-1}{\cal F}'''} \right|^2}. 
\ea
We also see that the mass scales of the problem are set by $m_{s*}$, 
the scalar gluon mass  and  $g^{-1} \overline{{\cal F}}'''_*$, the third prepotential derivative, (and $g^{-1} \partial g$), 
once the stationary value of the scalar is determined.

\subsection{Treatment of UV infinity}

In the NJL theory \cite{NJL1, NJL2}, there is only one coupling constant carrying dimension $-1$ and 
the dimensionless quantity is naturally formed by combining it with the relativistic cutoff, 
which is interpreted as the onset of UV physics. 
In the theory under our concern, 
UV physics is specified by the three input functions, $K, {\cal F}, W$ 
and the UV scales and infinities reside in some of the coefficients. 
Our supersymmetric counterterm \cite{imaru1, imaru2} is
\ba
V_{{\rm c.t.}} = -\frac{1}{2} {\rm Im} \int d^2\theta \Lambda {\cal W}^{0\alpha} {\cal W}_{0\alpha} 
= -\frac{1}{2} ({\rm Im} \Lambda) D^2. 
\ea
It is a counterterm associated with ${\rm Im}{\cal F}''$. 
We set up a renormalization condition
\ba
\left. \frac{1}{N^2} \frac{\partial^2 V}{(\partial D)^2}
 \right|_{D=0, \varphi = \varphi_*, \bar{\varphi} = \bar{\varphi}_* } =2c, 
\label{rencond}
\ea 
and relate (or transmute) the original infinity of the dimensional reduction scheme with that of ${\rm Im}{\cal F}''$. 
 We have indicated that this condition is set up at $D=0$ and the stationary point of the scalar which we will determine.
We stress again that the entire scheme is supersymmetric. 

\subsection{The one-loop part}

The entire contribution of all particles in the theory to $i \cdot$ (the 1PI to one-loop) $\equiv i \Gamma_{1-loop}$ 
is easy to compute, knowing (\ref{ev1}), (\ref{ev2}) and (\ref{sgluonmass}).
It is given by
\ba
i \Gamma_{{\rm 1-loop}} = 
\left(
\int d^4x
\right)
\sum_a \int \frac{d^4k}{(2\pi)^4} \ln 
\left(
\frac{(|\Lambda_a^{(+)}|^2 - k^2 - i \varepsilon)(|\Lambda_a^{(-)}|^2 - k^2 - i \varepsilon)}{(|m_{s,a}|^2 - k^2 - i \varepsilon)(-k^2 - i \varepsilon)}
\right). 
\label{1loopeffaction}
\ea
In the unbroken $U(N)$ phase, 
it is legitimate to replace $\displaystyle  \sum_a$ by $N^2$ and drop the index $a$ 
as we have said before.
We obtain 
\ba
V_{{\rm 1-loop}} &\equiv& (-i) \frac{1}{(\int d^4x)} \Gamma_{{\rm 1-loop}} \\
&=& -N^2 \left| {\rm tr}{\cal M} \right|^4 \int \frac{d^4l^\mu}{(2\pi)^4 i} 
\ln 
\left(
\frac{(|\lambda^{(+)}|^2 - l^2 - i \varepsilon)(|\lambda^{(-)}|^2 - l^2 - i \varepsilon)}{(\left| \frac{m_{s}}{{\rm tr}{\cal M}} \right|^2 - l^2 - i \varepsilon)(-l^2 - i \varepsilon)}
\right) \nonumber \\
&\equiv& N^2 |{\rm tr}{\cal M}|^4 J. 
\label{Jintroduced}
\ea
Note that $|m_s|^2$, whose stationary value give the tree mass squared of the scalar gluon, 
differ from $|{\rm tr}{\cal M}|^2$: 
\ba
\left| {\rm tr}{\cal M} \right|^2 = \left| m_s -\frac{i}{2} (g^{-1}{\cal F}''') F + (g^{-1}\partial g) \bar{F} \right|^2. 
\ea
To evaluate the integral in $d$-dimensions, we just quote
\ba
I(x^2) &\equiv& -\int \frac{d^4l^\mu}{(2\pi)^4 i} \log (x^2 - l^2 - i \varepsilon), \\
I(x^2)-I(0) &=& \frac{1}{32\pi^2} \left[ A(\varepsilon, \gamma) (x^2)^2 -(x^2)^2 \log (x^2) \right]
\ea
where
\ba
A(\varepsilon, \gamma) = \frac{1}{2} -\gamma +\frac{1}{\varepsilon}, \qquad \varepsilon = 2-\frac{d}{2}. 
\ea
We obtain
\ba
V_{{\rm 1-loop}} &=& \frac{N^2|{\rm tr}{\cal M}|^4}{32\pi^2} 
\left[
A(\varepsilon, \gamma) 
\left(
|\lambda^{(+)}|^4 + |\lambda^{(-)}|^4 - \left|\frac{m_s}{{\rm tr}{\cal M}} \right|^4
\right) \right. \nonumber \\
&& \left. 
-|\lambda^{(+)}|^4\log |\lambda^{(+)}|^2 
-|\lambda^{(-)}|^4\log |\lambda^{(-)}|^2
+\left| \frac{m_s}{{\rm tr}{\cal M}} \right|^4 \log \left| \frac{m_s}{{\rm tr}{\cal M}} \right|^4
\right].
\label{1loopeffpot}
\ea
This again depends upon $\Delta$, $f$ and $\varphi$.

\section{Stationary conditions and gap equation}
\subsection{Variational analyses}

Now we turn to our variational problem. 
It is stated as in the tree case as
\ba
&&\frac{\partial V}{\partial D} = 0, \label{Dflat} \\
&&\frac{\partial V}{\partial F} = \frac{\partial V}{\partial \bar{F}} = 0, \label{Fflat}\\
&&\frac{\partial V}{\partial \varphi} = \frac{\partial V}{\partial \bar{\varphi}} = 0. \label{phiflat}
\ea
We will regard the solution to be obtained by considering Eqs. (\ref{Dflat}) and (\ref{phiflat}) first and 
solving $D$ and $\varphi$ for $F$ and $\bar{F}$:
\ba
D=D_*(F, \bar{F}), \quad \varphi = \varphi_*(F, \bar{F}), \quad \bar{\varphi} = \bar{\varphi}_*(F, \bar{F}).
\ea
Eq. (\ref{Fflat}) is then
\ba
\left. 
\frac{\partial V(D=D_*(F, \bar{F}), \varphi = \varphi_*(F, \bar{F}), \bar{\varphi} = \bar{\varphi}_*(F, \bar{F}), F, \bar{F})}{\partial F}
\right|_{D, \varphi, \bar{\varphi}, \bar{F}~{\rm fixed}} =0
\ea
and its complex conjugate. 
These will determine $F=F_*, \bar{F}=\bar{F}_*$. 

In this paper, we are going to work in the region
where the magnitude $\left| F_* \right|$ is small and can be treated perturbatively. 
This means that, in the leading order, the problem posed by Eq. (\ref{Dflat}) and Eq. (\ref{phiflat}) becomes
\ba
&&\frac{\partial V(D, \varphi, \bar{\varphi}, F=0, \bar{F}=0)}{\partial D} = 0, \label{Dflatleading} \\
&&\frac{\partial V(D, \varphi, \bar{\varphi}, F=0, \bar{F}=0)}{\partial \varphi} =  \frac{\partial V(D, \varphi, \bar{\varphi}, F=0, \bar{F}=0)}{\partial \bar{\varphi}} = 0.
\label{phiflatleading}
\ea
Eq. (\ref{Dflatleading}) is nothing but the gap equation given in \cite{imaru1, imaru2}, 
while Eq. (\ref{phiflatleading}) is the stationary conditions for the scalar.  
This is the variational problem which we should solve. 
A set of stationary values $(D_*, \varphi_*, \bar{\varphi}_*)$ is determined as the solution.

\subsection{The analysis in the region $F_* \approx 0$}

Let us first determine $V(D, \varphi, \bar{\varphi}, F=0, \bar{F}=0)$ explicitly. 
We need to solve the normalization condition. 
\ba
2c N^2 = \left. \frac{\partial^2 V}{(\partial D)^2} \right|_{D=0,_*} 
= -({\rm Im} {\cal F}''_*) - ({\rm Im} \Lambda) + N^2 |{\rm tr}{\cal M}|^4 
\left. \frac{\partial^2 J}{(\partial D)^2} \right|_{D=0},
\ea
where $J$ has been introduced in Eq.~(\ref{Jintroduced}).
At $F, \bar{F} \to 0$, 
\ba
\Delta &\to& \Delta_0 \equiv r_0(\varphi, \bar{\varphi}) D, \quad \quad
r_0 = \frac{\sqrt{2}}{2} \frac{\sqrt{g^{-1}({\rm Im}{\cal F''})^{-1}}{\cal F}'''}{g^{-1}W''}, \\
\lambda^{(\pm)} &\to& \lambda^{(\pm)}_0 = \frac{1}{2} \left( 1\pm \sqrt{1 + \Delta_0^2} \right), 
\ea
where
\ba 
&&\frac{m_s}{{\rm tr}{\cal M}} \to 1, \\
&&J \to J_0 \equiv \frac{1}{32\pi^2} 
\left[
A(\varepsilon, \gamma) \left\{ \frac{1}{2} \left( 1+\frac{1}{2}\Delta_0^2 \right)\left( 1+\frac{1}{2} \bar{\Delta}_0^2\right) 
+ \frac{1}{2}\sqrt{1+ \Delta_0^2} \sqrt{1 + \bar{\Delta}_0^2} -1
\right\} \right. \nonumber \\
&& \left. \hspace*{3.5cm} -|\lambda_0^{(+)}|^4 \log |\lambda_0^{(+)}|^2 - |\lambda_0^{(-)}|^4 \log |\lambda_0^{(-)}|^2
\right],
\ea
essentially reducing the situation to that of Refs. \cite{imaru1, imaru2}. 

Note, however, that $r$ and $\Delta~({\rm or}~r_0, \Delta_0)$ are complex in general except those special cases 
which include the case of the rigid ${\cal N}=2$ supersymmetry partially broken to ${\cal N}=1$ at the tree vacua. 
For $|\Delta_0| \ll 1$, 
\ba
J_0 \approx \frac{1}{32\pi^2} 
\left[
A(\varepsilon, \gamma) \frac{1}{2}(\Delta_0^2 + \bar{\Delta}_0^2) -\frac{1}{4} (\Delta_0^2 + \bar{\Delta}_0^2) + {\cal O}(|\Delta_0|^{4-\varepsilon}) 
\right]. 
\ea
We solve the normalization condition for the number $A$ to obtain
\ba
A = \frac{1}{2} + \frac{32\pi^2}{|m_{s*}|^4(r_{0*}^2+\bar{r}_{0*}^2)} \left (2c + \frac{{\rm Im}{\cal F}''_*}{N^2} + \frac{{\rm Im}\Lambda}{N^2} \right) 
\equiv \tilde{A}(c, \Lambda, \varphi_*, \bar{\varphi}_*). 
\ea
We obtain
\ba
V_0 &=& V(D, \varphi, \bar{\varphi}, F=0, \bar{F}=0) \nonumber \\
&=& -\frac{1}{2} {\rm Im}{\cal F}'' D^2 -\frac{1}{2} ({\rm Im}\Lambda) D^2 \nonumber \\
&&+ \frac{N^2|m_s|^4}{32\pi^2} 
\left[
\tilde{A}(c, \Lambda, \varphi_*, \bar{\varphi}_*) 
\left\{
\frac{1}{2} \left( 1+ \frac{1}{2} \Delta_0^2 \right) \left( 1+ \frac{1}{2} \bar{\Delta}_0^2 \right) + \frac{1}{2} \sqrt{1+\Delta_0^2} \sqrt{1+\bar{\Delta}_0^2} -1
\right\} \right. \nonumber \\
&& \left. -|\lambda_0^{(+)}|^4 \log |\lambda_0^{(+)}|^2 - |\lambda_0^{(-)}|^4 \log |\lambda_0^{(-)}|^2
\right]. 
\label{renpotential}
\ea
After some calculation, this is found to be expressible as
\ba
\frac{V_0}{N^2 |m_s|^4} &=& 
\left(
\frac{1}{64\pi^2} + \tilde{c} - \tilde{\delta}(\varphi, \bar{\varphi})
\right)
\left( \frac{\Delta_0 + \bar{\Delta}_0}{2} \right)^2 
+ \frac{1}{32\pi^2} \tilde{A} \left( \frac{1}{8} |\Delta_0|^4 + f(\Delta_0, \bar{\Delta}_0) \right) \nonumber \\
&&-\frac{1}{32\pi^2} \left( |\lambda_0^{(+)}|^4 \log |\lambda_0^{(+)}|^2 + |\lambda_0^{(-)}|^4 \log |\lambda_0^{(-)}|^2 \right), 
\label{renpotential2}
\ea
where
\ba
\tilde{c} &=& \frac{c}{|m_{s*}|^4 \left( \frac{r_{0*}^2 + \bar{r}_{0*}^2}{2} \right)}, \\
\tilde{\delta}(\varphi, \bar{\varphi}) &=& \frac{1}{2} 
\left(
\frac{ \frac{{\rm Im}{\cal F}_*''}{N^2} + \frac{{\rm Im}\Lambda}{N^2} }{ \frac{ (r_{0*}^2 + \bar{r}_{0*}^2) }{2} |m_{s*}|^4} 
\right)
\left[
\frac{\frac{{\rm Im}{\cal F}''/N^2 + {\rm Im} \Lambda/N^2}{{\rm Im}{\cal F}_*''/N^2 
+ {\rm Im}\Lambda/N^2} }{\frac{|m_s|^4}{|m_{s*}|^4} \frac{\left(\frac{r_0+\bar{r}_0}{2} \right)^2}{\left( \frac{r_{0*}^2 + \bar{r}_{0*}^2}{2} \right) } } 
- 1
\right], 
\\
f(\Delta_0, \bar{\Delta}_0) &=& \frac{1}{2} \left( \sqrt{1+\Delta_0^2} \sqrt{1+\bar{\Delta}_0^2} - |\Delta_0|^2 -1 \right).
\ea
Note that 
\ba
\tilde{\delta}_* \equiv \tilde{\delta}(\varphi_*, \bar{\varphi}_*) \ne 0,
\ea
and
\ba
\left| f(\Delta_0, \bar{\Delta}_0) \right| \le {\rm const} \quad {\rm for}~\left| \Delta_0 \right| \gg 1. 
\ea

If $r_0$ (and $\Delta_0$) is real, this is rewritten as
\ba
\frac{V_0}{N^2|m_s|^4} = \left( \left(c' + \frac{1}{64\pi^2} \right) -\delta \right) \Delta_0^2 
+ \frac{1}{32\pi^2} 
\left[
\frac{\tilde{A}}{8} \Delta_0^4 
- {\lambda_0^{(+)}}^4 \log {\lambda_0^{(+)}}^2 - {\lambda_0^{(-)}}^4 \log {\lambda_0^{(-)}}^2
\right],  
\label{realpotential}
\ea
where $c' \equiv \frac{c}{r_{0*}^2 |m_{s*}|^4}$ is the rescaled number, and 
\ba
\delta(\varphi, \bar{\varphi}) \equiv \frac{1}{2} 
\left(
\frac{\frac{{\rm Im}{\cal F}''_*}{N^2} + \frac{{\rm Im}\Lambda}{N^2}}{r_{0*}^2 |m_{s*}|^4}
\right)
\left[
\frac{\frac{{\rm Im}{\cal F}''/N^2+ {\rm Im}\Lambda/N^2}{{\rm Im}{\cal F}''_*/N^2 + {\rm Im}\Lambda/N^2}}{\frac{r_0^2|m_s|^4}{r_{0*}^2 |m_{s*}|^4}}
-1
\right]. 
\label{delta}
\ea
Clearly, there are two scales in our current problem $|r_{0*}|^{-1/2}$ and $|m_{s*}|$, 
which are controlled by the second superpotential derivative and the third prepotential derivative at the stationary value $\varphi_*$. 

Let us turn to the gap equation 
\ba
\left. \frac{\partial V_0}{\partial D} \right|_{\varphi, \bar{\varphi}} = 0. 
\ea
For Eq. (\ref{renpotential2}), scaling out $|r_0|^2$, we find
\ba
0 &=& D
\left[
\left( \frac{1}{64\pi^2} + \tilde{c} - \tilde{\delta} \right) (1 + \cos 2 \theta) 
+ \frac{\tilde{A}}{32\pi^2} 
\left\{
\frac{1}{2}|\Delta_0|^2 - (1 - \cos 2 (\theta - \theta'))
\right\} \right. \nonumber \\
&& \left. 
-\frac{1}{32\pi^2} 
\left\{
\left( 2 \log |\lambda_0^{(+)}|^2 + 1 \right) \frac{1}{2} 
\left(
\frac{e^{2i\theta}\bar{\lambda}_0^{(+)}}{\sqrt{1+\Delta_0^2}} + \frac{e^{-2i\theta}\lambda_0^{(+)}}{\sqrt{1+\bar{\Delta}_0^2}}
\right) |\lambda_0^{(+)}|^2 
\right. \right. \nonumber \\
&& \left. \left. 
- \left( 2 \log |\lambda_0^{(-)}|^2 + 1 \right) \frac{1}{2} 
\left(
\frac{e^{2i\theta}\bar{\lambda}_0^{(-)}}{\sqrt{1+\Delta_0^2}} + \frac{e^{-2i\theta}\lambda_0^{(-)}}{\sqrt{1+\bar{\Delta}_0^2}}
\right) |\lambda_0^{(-)}|^2
\right\}
\right], 
\label{gap1}
\ea
where 
\ba
\Delta_0 = |\Delta_0| e^{i\theta}, \quad r_0 = |r_0| e^{\i\theta}, \quad \tan 2\theta' = \frac{|\Delta_0|^2 \sin 2\theta}{1+ |\Delta_0|^2 \cos 2\theta}. 
\ea
Note that $|1-\cos 2(\theta-\theta')| \to 0$ in the region $\theta \sim 0$ or $|\Delta_0| \gg 1$. 

On the other hand, for Eq. (\ref{realpotential}) with $\Delta_0$ being real, $N^2 |m_s|^4$ is scaled out and 
it is simply given by the $\Delta_0$ derivative:
\ba
0 &=&
\Delta_0 
\left[
2 \left( 
\left( c' + \frac{1}{64 \pi^2} \right) - \delta 
\right) \right. \nonumber \\ 
&& \left. 
+ \frac{1}{32\pi^2}
\left\{
\frac{\tilde{A}}{2} \Delta_0^2 
-\frac{1}{\sqrt{1 + \Delta_0^2}} 
\left(
\lambda_0^{(+)3} \left( 2 \log \lambda_0^{(+)2} + 1 \right)
- \lambda_0^{(-)3} \left( 2 \log \lambda_0^{(-)2} + 1 \right)
\right)
\right\}
\right], \nonumber \\
\label{gap2}
\ea
which is our original gap equation.\footnote{We have introduced $\delta(\varphi, \bar{\varphi})$ 
such that its stationary value $\delta(\varphi_*, \bar{\varphi}_*)=0$, which can therefore be ignored in
 analyzing Eq. (\ref{gap2}).}
In both cases, the solutions are given by the extremum of the potential $V_0(D, \varphi, \bar{\varphi})$ in its $D$ profile. 
We stress again that the $D$ profile is not a direct stability criterion of the vacua, 
which is to be discussed with regard to the scalar potential $V_0(D_*(\varphi, \bar{\varphi}), \varphi, \bar{\varphi})$.

We next examine $\left. \frac{\partial V_0}{\partial \varphi} \right|_{D, \bar{\varphi}}=0$ and its complex conjugate. 
For Eq. (\ref{renpotential2}), we obtain
\ba
2 \frac{\partial}{\partial \varphi} (\ln |m_s|^2) \frac{V_0}{N^2 |m_s|^4} &=& \left( \frac{\partial \tilde{\delta}}{\partial \varphi} \right) 
\left( \frac{\Delta_0 + \bar{\Delta}_0}{2} \right)^2 
-D \hat{{\cal P}} \left( \frac{V_0}{N^2 |m_s|^4} \right) \nonumber \\
&& -D \left[ \left. \frac{\partial \ln r_0}{\partial \varphi} \right|_{\bar{\varphi}} + \left. \frac{\partial \ln \bar{r}_0}{\partial \bar{\varphi}} \right|_{\varphi} \right]
\frac{\partial}{\partial D} \left( \frac{V_0}{N^2 |m_s|^4} \right), 
\label{phiflat1}
\ea
and its complex conjugate
where 
\ba
\hat{{\cal P}} = i \left( \frac{\partial \theta}{\partial \varphi} \right) 
\left(
r_0 \frac{\partial}{\partial \Delta_0} - \bar{r}_0 \frac{\partial}{\partial \bar{\Delta}_0}
\right). 
\ea
The second term of the RHS of Eq. (\ref{phiflat1}) is proportional to the gap equation Eq. (\ref{gap1}). 
As for the third term,  after some calculation, we obtain
\ba
\frac{-\hat{{\cal P}}\left( \frac{V_0}{N^2 |m_s|^4} \right)}{\left(\frac{\partial \theta}{\partial \varphi} \right)|\Delta||r_0|} 
&=& \left( \frac{1}{64\pi^2} + \tilde{c} - \tilde{\delta} \right) \sin 2 \theta 
+ \frac{1}{32\pi^2} \tilde{A} \sin 2(\theta-\theta') \nonumber \\
&&- \frac{1}{32\pi^2} \frac{1}{2} \left( \frac{\sin(2\theta-\theta')}{|1+\Delta_0^2|^{1/2}} + \sin 2 (\theta - \theta') \right) 
|\lambda_0^{(+)}|^2 \left( 2 \log|\lambda_0^{(+)}|^2 +1 \right) \nonumber \\
&&+ \frac{1}{32\pi^2} \frac{1}{2} \left( \frac{\sin(2\theta-\theta')}{|1+\Delta_0^2|^{1/2}} - \sin 2 (\theta - \theta') \right) 
|\lambda_0^{(-)}|^2 \left( 2 \log |\lambda_0^{(-)}|^2 +1 \right) \nonumber \\
&&\equiv C(\theta, |\Delta_0|). 
\label{phiflat11}
\ea
In the RHS of Eqs. (\ref{phiflat1}) and (\ref{phiflat11}), 
we have regarded $\Delta_0, \bar{\Delta}_0, \varphi$ and $\bar{\varphi}$ as independent variables. 

For Eq. (\ref{realpotential}), with $\Delta_0$ real, we obtain
\ba
2 \partial(\ln |m_s|^2) \frac{V_0}{N^2|m_s|^4} = \left( \frac{\partial \delta}{\partial \varphi} \right) \Delta_0^2 
-\frac{\partial \Delta_0}{\partial \varphi} \frac{\partial}{\partial \Delta_0} \left(\frac{V_0}{N^2 |m_s|^4} \right)
\label{phiflat2}
\ea
and its complex conjugate. 
Here in the last term of the RHS, we have regarded $\Delta_0, \varphi, \bar{\varphi}$ as independent variables. 
\begin{figure}[htbp]
 \begin{center}
  \includegraphics[width=60mm]{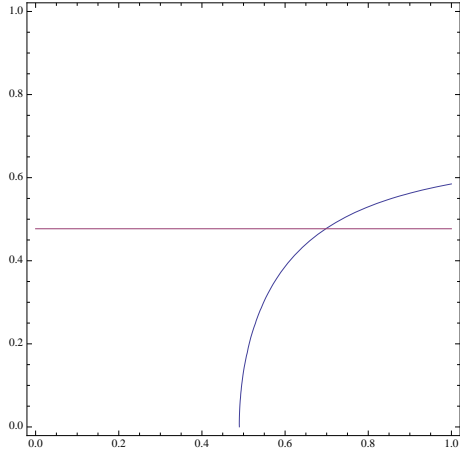}
 \end{center}
 \caption{The schematic picture of the intersection of the two curves which represent 
 the solution to the gap equation (the red one) 
 and the $\varphi$ flat condition (the blue one). The horizontal axis is denoted by $\varphi/M$ and 
the vertical one by $\Delta_0$. 
 The values at the stationary point ($\Delta_{0*}, \varphi_*=\bar{\varphi}_*$) are
  read off from the intersection point. }
 \label{intersection}
\end{figure}

Finally the stationary values ($D_*, \varphi_*, \bar{\varphi}_*$) are determined by Eqs. (\ref{gap1}) and 
(\ref{phiflat1}) or by Eqs. (\ref{gap2}) and (\ref{phiflat2}).  
Let us discuss the latter case first. 
As the second term of the RHS in Eq. (\ref{phiflat2}) is nothing but
 the gap equation Eq. (\ref{gap2}), Eq. (\ref{phiflat2}) can be safely replaced by 
\ba
\frac{V_0}{N^2|m_s|^4} = \frac{\frac{\partial \delta}{\partial \varphi}}{2 \partial (\ln |m_s|^2)} \Delta_0^2,
\label{7}
\ea
$\varphi$ being real. 
The solution to Eq. (\ref{7}) in the $\Delta_0$ profile is determined as the point of intersection of the potential 
with the quadratic term having $\varphi=\bar{\varphi}$ dependent coefficients. 
Actually, it is a real curve in the full ($\Delta_0, \varphi=\bar{\varphi}$) plane. 
Likewise, the solution to the gap equation Eq. (\ref{gap2}), the condition of $\Delta_0$ extremum of the potential, 
provides us with another real curve in the  ($\Delta_0, \varphi=\bar{\varphi}$) plane. 
The values ($\Delta_{0*}, \varphi_*=\bar{\varphi}_*$) are the intersection of these two. 
The schematic figure of the intersection is displayed in Figure \ref{intersection}. 
By tuning our original input functions, it is possible to arrange such intersection. 
Conversely, as an inverse problem, for given $\Delta_{0*}$ and the height of the $\Delta_0$ profile, 
one can always find the values of the coefficients in Eq. (\ref{gap2}) and the coefficient function in Eq. (\ref{7}) 
that accomplish this. 
Dynamical supersymmetry breaking has been realized. 

As for the former case, as in the latter case, we can safely replace Eq. (\ref{phiflat1}) by 
\ba
\frac{V_0}{N^2 |m_s|^4} = \frac{\left( \frac{\partial \tilde{\delta}}{\partial \varphi} \right)}{2\partial (\ln |m_s|^2)} 
\left( \frac{\Delta_0 + \bar{\Delta}_0}{2} \right)^2 
+ \frac{\left( \frac{\partial \theta}{\partial \varphi} \right) |\Delta_0|^2}{2\partial(\ln|m_s|^2)}C.
\label{phiflat1final}
\ea
The values ($\Delta_{0*}, \bar{\Delta}_{0*}, \varphi_*, \bar{\varphi}_*$) can be determined 
by the intersection of Eq. (\ref{gap1}) and Eq. (\ref{phiflat1final}). 
We will not carry out the (numerical) analysis for this case further in this paper.

\subsection{Determination of $F_*$}
Let us now turn to the analysis of the remaining equation of our variational problem, Eq. (\ref{Fflat}). 
In our current treatment, 
\ba
F = -\frac{1}{g} \overline{W}' + \frac{1}{g} \frac{\partial}{\partial \bar{F}} V_{{\rm 1-loop}} \approx 
\left. -\frac{1}{g} \overline{W}' + \frac{1}{g} \frac{\partial}{\partial \bar{F}} V_{{\rm 1-loop}} \right|_{F=0}.
\ea
As the stationary values $(D_*, \varphi_*, \bar{\varphi}_*)$ are already determined, 
this equation and its complex conjugate determine $F_*$ and $\bar{F}_*$:
\ba
F_* = \frac{1}{g(\varphi_*, \bar{\varphi}_*)} \left( \left. -\overline{W}'(\varphi_*, \bar{\varphi}_*) 
+ \frac{\partial}{\partial \bar{F}} V_{{\rm 1-loop}}(D_*, \varphi_*, \bar{\varphi}_*, F, \bar{F}) \right|_{F=\bar{F}=0} \right).
\ea
Note that, knowing $V_{{\rm 1-loop}}$ explicitly in Eq. (\ref{1loopeffpot}), 
the RHS can be evaluated. 
We can check the consistency of our treatment through $f_3$ in Eq. (\ref{f3}) by $|f_3| \ll 1$.

\subsection{Numerical study of the gap equation}

In this subsection, we study some numerical solutions to the gap equation Eq. (\ref{gap2}) 
and the stationary condition for $\varphi$ Eq. (\ref{7}) in the real $\Delta_0$ case. 
The equations we should study are
\ba
&&0 = 2 
\left( c' + \frac{1}{64 \pi^2} \right) 
+ \frac{1}{32\pi^2}
\left\{
\frac{\tilde{A}}{2} \Delta_{0*}^2 
\right. \nonumber \\
&& \left. \left. 
-\frac{1}{\sqrt{1 + \Delta_{0*}^2}} 
\left(
\lambda_0^{(+)3} \left( 2 \log \lambda_0^{(+)2} + 1 \right)
- \lambda_0^{(-)3} \left( 2 \log \lambda_0^{(-)2} + 1 \right)
\right) \right|_{\Delta_0=\Delta_{0*}}
\right\}, 
\label{gapatsp} \\
&&\frac{V_0}{N^2|m_{s*}|^4} = \frac{\left. \frac{\partial \delta(\varphi, \bar{\varphi})}{\partial \varphi} \right|_{\varphi_*, \bar{\varphi}_*}}{2 \partial (\ln |m_{s*}|^2)} \Delta_{0*}^2, 
\ea
where we note that $\delta(\varphi, \bar{\varphi})$ in the gap equation (\ref{gap2}) vanishes at the stationary point in the real $\Delta_0$ case.  
By using Eqs. (\ref{realpotential}) and (\ref{delta}), the second condition can be rewritten after dividing by $\Delta_{0*}^2$ as
\ba
&&\left(c' + \frac{1}{64\pi^2} \right) 
+ \frac{1}{32\pi^2} 
\left[
\frac{\tilde{A}}{8} \Delta_{0*}^2 
- \left. \frac{1}{\Delta_{0*}^2}\left(  |\lambda_0^{(+)}|^4 \log |\lambda_0^{(+)}|^2 - |\lambda_0^{(-)}|^4 \log |\lambda_0^{(-)}|^2 \right) \right|_{\Delta_0=\Delta_{0*}}
\right] \nonumber \\
&=& \frac{1}{4N^2 \partial \ln |m_{s*}|^2} \frac{{\rm Im}({\cal F}_*'' + \Lambda)}{r_{0*}^2 | m_{s*} |^4} 
\left[
\partial \left. \ln {\rm Im}({\cal F}''+\Lambda) \right|_{\varphi_*, \bar{\varphi}_*} 
- \frac{\partial (\left. r_0 |m_s|^2)^2 \right|_*}{(r_{0*}^2 |m_{s*}|^4)^2}
\right]. 
\label{phiflatatsp}
\ea
The nontrivial solution $\Delta_{0*} \ne 0$ to the gap equation (\ref{gapatsp}) is found by some region of the
 parameters $c'$ and $\tilde{A}$, which was already done in \cite{imaru1}. 
This solution fixes the LHS of Eq. (\ref{phiflatatsp}) and $\varphi_*$ is determined by solving Eq. (\ref{phiflatatsp}) in principle. 
In order to find $\varphi_*$ explicitly, the form of the prepotential ${\cal F}$ and  that of the superpotential $W$ 
must be specified. 
Here, we take a simple prepotential and a superpotential of the following type 
\ba
&&{\cal F} = \frac{c}{2N}  {\rm tr} \varphi^2 + \frac{1}{3!MN} {\rm tr} \varphi^3  \equiv \frac{1}{2} c \varphi^2 + \frac{1}{3!M} \varphi^3, 
\label{F} \\
&&W = \frac{m^2}{N} {\rm tr} \varphi + \frac{d}{3!N} {\rm tr} \varphi^3 \equiv m^2 \varphi + \frac{d}{3!} \varphi^3,
\label{W}
\ea
where $c, d$ are dimensionless constants while $m, M$ are dimensionful parameters. 
In particular, $M$ is a cutoff scale of the theory.  
This prepotential is minimal for DDSB. 
As for the superpotential, 
at least two terms are required to be supersymmetric at tree level. 
We can take a quadratic term $\varphi^2$ instead of the cubic one, 
but in that case, RHS of Eq. (\ref{phiflatatsp}) becomes singular because of $\partial \ln |m_s|^2=0$. 

Substituting these ${\cal F}$ and $W$ into Eq. (\ref{phiflatatsp}), we obtain
\ba
&&\left(c' + \frac{1}{64\pi^2} \right) 
+ \frac{1}{32\pi^2} 
\left[
\frac{\tilde{A}}{8} \Delta_{0*}^2 
- \left. \frac{1}{\Delta_{0*}^2}\left(  |\lambda_0^{(+)}|^4 \log |\lambda_0^{(+)}|^2 - |\lambda_0^{(-)}|^4 \log |\lambda_0^{(-)}|^2 \right) \right|_{\Delta_0=\Delta_{0*}}
\right] \nonumber \\
&=& -\frac{{\rm Im}(c + \Lambda)({\rm Im}~c)^4}{N^2} \frac{1}{\left( d \varphi_*/M \right)^2},
\label{phiflatfinal}
\ea
where we utilized the fact that $1/M, d, \varphi_*$ are real and $c$ is pure imaginary, which are necessary for $\Delta_0=\bar{\Delta}_0$. 
Taking the coefficients $c=i, d=1$ for further simplification, 
we can easily obtain a solution by tuning $N$ and ${\rm Im}\Lambda$.  
We note $0 \le \varphi_*/M \le 1$ for our effective theory to be valid. 
In our analysis carried out in this paper, we consider the region where the magnitude of the F-term is smaller compared to that of the D-term. 
Therefore, we need to check whether our solutions satisfy this property consistently. 
Let us consider the ratio of the auxiliary fields:  
\ba
\left| \frac{F_*}{D_*} \right| &=& \left| \frac{-g^{-1} \overline{W}'(\bar{\varphi}_*) 
+ g^{-1} \left. \frac{\partial}{\partial \bar{F}} V_{{\rm 1-loop}}(D_*, \varphi_*, \bar{\varphi}_*, F, \bar{F}) \right|_{F=\bar{F}=0}}{\Delta_{0*}/r_{0*}} \right| \nonumber \\
&=& \left| 
\frac{1}{\sqrt{2}\Delta_{0*} \frac{\varphi_*}{M}} 
\left[
\left( \frac{m}{M} \right)^2 + \frac{1}{2} \left( \frac{\varphi_*}{M} \right)^2
\right] 
\right. \nonumber \\
&& \left. 
+ i \frac{N^2}{\sqrt{2}\Delta_{0*}} \left( \frac{\varphi_*}{M} \right)^2
\left[
\frac{\tilde{A}}{128\pi^2} \Delta_{0*}^2
-\frac{1}{32\pi^2} (|\lambda_{0*}^+|^4 \log |\lambda_{0*}^+|^2 + |\lambda_{0*}^-|^4 \log |\lambda_{0*}^-|^2 +1) 
\right. \right. \nonumber \\
&& \left. \left. 
+\frac{1+ \frac{\Delta_{0*}^2}{2}}{32\pi^2 \sqrt{1+\Delta_{0*}^2}} 
\left\{
(\lambda_{0*}^+)^3 \left(\log |\lambda_{0*}^+|^2 + \frac{1}{2} \right) - (\lambda_{0*}^-)^3 \left(\log |\lambda_{0*}^-|^2 + \frac{1}{2} \right) 
\right\} 
\right]
\right|,
\ea
where the form of the prepotential and that of the superpotential in Eq. (\ref{F}) and Eq. (\ref{W}) are assumed
 and we have put $c=i, d=1$ in the second equality. 

Now, the numerical solutions to the gap equation and the stationary condition for $\varphi$ are listed in Table \ref{numerical}. 
In these examples, we have taken some values of $-\frac{N^2}{{\rm Im}(i + \Lambda)}$ and $m$ just for an illustration and 
the ratio $|F_*/D_*|$ and $|f_{3*}|$ are evaluated. 
We can find that the $F$-term is smaller than the $D$-term in some of these examples. 
\begin{table}[htb]
\begin{center}
  \begin{tabular}{|c|c|c|c|c|c|} 
  \hline
$c'+\frac{1}{64\pi^2}$ & $\tilde{A}/(4\cdot32\pi^2)$ & $\Delta_{0*}$  
& $\varphi_*/M~(-\frac{N^2}{{\rm Im}(i+\Lambda)})$ & $|F_*/D_*|$ & $|f_{3*}|$ \\ 
\hline
0.002 & 0.0001 & 0.477 & 0.707~(10000) & 2.621~($m=M$) & 1.77 \\
0.002 & 0.0001 & 0.477 & 0.707~(10000) & 0.524~($m \ll M$) & 0.35 \\
0.002 & 0.0001 & 0.477 & 0.707~(10000) & 0.860~($m=0.4M$) & 0.58 \\
\hline
0.003 & 0.001 & 1.3623 & 0.8639~(2000) & 0.825~($m=M$) & $>$1 \\
0.003 & 0.001 & 1.3623 & 0.8639~(2000) & 0.224~($m \ll M$) & 0.43 \\
0.003 & 0.001 & 1.3623 & 0.5464~(5000) & 1.092~($m=M$) & $>$1 \\
0.003 & 0.001 & 1.3623 & 0.5464~(5000) & 0.142~($m \ll M$) & 0.27 \\
0.003 & 0.001 & 1.3623 & 0.5464~(5000) & 0.911~($m=0.9M$) & 1.76 \\
0.003 & 0.001 & 1.3623 & 0.3863~(10000) & 1.444~($m=M$) & $>$1 \\
0.003 & 0.001 & 1.3623 & 0.3863~(10000) & 0.100~($m \ll M$) & 0.19 \\
0.003 & 0.001 & 1.3623 & 0.3863~(10000) & 0.960~($m=0.8M$) & 1.85 \\
 \hline
  \end{tabular}
\end{center}
\caption{Samples of numerical solutions for the gap equation and the stationary condition for $\varphi$. 
The ratio $|F_*/D_*|$ and $|f_{3*}|$ are also evaluated for consistency check.}
\label{numerical}
\end{table}

\subsection{
Mass of the scalar gluons}

We now turn to the question of the second variations of the scalar
\ba
V_{{\rm scalar}} = V(D=D_*(\varphi, \bar{\varphi}), F=F_*(\varphi, \bar{\varphi}) \approx 0, \bar{F}=\bar{F}_*(\varphi, \bar{\varphi}) \approx 0, \varphi, \bar{\varphi}) 
\ea
at the stationary point ($D_*(\varphi_*,\bar{\varphi}_*), 0, 0, \varphi_*, \bar{\varphi}_*$).\footnote{The discussion of this subsection has been further updated in \cite{imaru5}.} 
It is convenient to separate $V(D, F, \bar{F}, \varphi, \bar{\varphi})$ into two parts: 
\ba
V = {\cal V}+V_0. 
\ea
Here
\ba
{\cal V}(F, \bar{F}, \varphi, \bar{\varphi}) &\approx& - g F \bar{F} - F W' - \bar{F} \bar{W}' + (\partial_F V_{{\rm 1-loop}})_* F + (\partial_{\bar{F}} V_{{\rm 1-loop}})_* \bar{F} \nonumber \\
&&+ \frac{1}{2} (\partial_F^2 V_{{\rm 1-loop}})_* F^2 + \frac{1}{2} (\partial^2_{\bar{F}} V_{{\rm 1-loop}})_* \bar{F}^2 + (\partial_F \partial_{\bar{F}} V_{{\rm 1-loop}})_* F \bar{F},
\label{calV}
\ea
and 
\ba
V_0(D, \varphi, \bar{\varphi}) = V(D, \varphi, \bar{\varphi}, F=0, \bar{F}=0). 
\label{V0}
\ea
In Eq. (\ref{calV}), we have extracted the $F, \bar{F}$ dependence of $V_{{\rm 1-loop}}$ (Eq. (\ref{1loopeffpot})) 
and $_*$ indicates that they are evaluated at $(D_*, \varphi_*, \bar{\varphi}_*, 0, 0)$ after the derivatives are taken. 
Eq. (\ref{V0}) has been computed in Eq. (\ref{renpotential}) and Eq. (\ref{realpotential}). 
We will compute the second partial derivatives and the second variations of $V_{{\rm scalar}}$. 

For ${\cal V}$, $\vec{y}_L = (F, \bar{F}), \vec{y}_R=(\varphi, \bar{\varphi})$, 
\ba
&&M_{RR_*} \equiv 
\left(
\begin{array}{cc}
\partial^2 {\cal V}, & \partial \bar{\partial} {\cal V} \\
\bar{\partial} \partial {\cal V}, & \bar{\partial}^2 {\cal V} \\
\end{array}
\right)_* \approx 0, \\
&&M_{RL_*} \equiv 
\left(
\begin{array}{cc}
\partial \partial_F {\cal V}, & \partial \partial_{\bar{F}} {\cal V} \\
\bar{\partial} \partial_F {\cal V}, & \bar{\partial} \partial_{\bar{F}} {\cal V} \\
\end{array}
\right)_* \approx 
\left(
\begin{array}{cc}
-W''+ (\partial \partial_F V_{{\rm 1-loop}}), & (\partial \partial_{\bar{F}} V_{{\rm 1-loop}}) \\
(\bar{\partial} \partial_F V_{{\rm 1-loop}}), & -\overline{W}'' + (\bar{\partial}\partial_{\bar{F}} V_{{\rm 1-loop}}) \\
\end{array}
\right)_*, \\
&&M_{LR_*} = M_{RL_*}^t, \\
&&M_{LL_*} \equiv 
\left(
\begin{array}{cc}
\partial_F^2 {\cal V}, & \partial_F \partial_{\bar{F}} {\cal V} \\
\partial_{\bar{F}} \partial_F {\cal V}, & \partial_{\bar{F}}^2 {\cal V} \\
\end{array}
\right)_* \approx 
\left(
\begin{array}{cc}
(\partial_F^2 V_{{\rm 1-loop}}), & -g + (\partial_F \partial_{\bar{F}} V_{{\rm 1-loop}}) \\
-g + (\partial_{\bar{F}} \partial_F V_{{\rm 1-loop}}), & (\partial_{\bar{F}}^2 V_{{\rm 1-loop}}) \\
\end{array}
\right)_*.
\ea

We obtain, after some computation, 
\ba
\delta^2 {\cal V}_* \approx \frac{1}{2} \delta \vec{y}_R^t M_{RL_*} (-M_{LL_*}^{-1}) M_{LR_*} \delta \vec{y}_R 
\equiv \frac{1}{2} \delta \vec{y}_R^{\dagger}
\left(
\begin{array}{cc}
{\cal M}_{\varphi\bar{\varphi}} & {\cal M}_{\varphi\varphi} \\
{\cal M}_{\bar{\varphi}\bar{\varphi}} & {\cal M}_{\bar{\varphi}\varphi} \\
\end{array}
\right)_* \delta \vec{y}_R 
\label{massmatrix}
\ea
where
\ba
&&{\cal M}_{\varphi\bar{\varphi}} = \frac{1}{g(G^2 - |C|^2)} \left( G (|A|^2 + |B|^2) + C A \bar{B} + \bar{C} \bar{A} B \right), \\
&&{\cal M}_{\varphi \varphi} = \frac{1}{g(G^2 - |C|^2)} \left( 2G A B + C A^2 + \bar{C} B^2 \right),
~G \equiv 1-\frac{\partial_F \partial_{\bar{F}}V_{{\rm 1-loop}}}{g}, \nonumber \\
&&A \equiv W'' -\partial \partial_F V_{{\rm 1-loop}},~
B \equiv -\partial \partial_{\bar{F}} V_{{\rm 1-loop}},~C \equiv \frac{\partial_{\bar{F}}^2 V_{{\rm 1-loop}}}{g}.
\ea
We see that in the region $|(\partial_F \partial_{\bar{F}} V)_0|_*, |(\partial_F^2 V)_0|_*, \ll g_*$, 
the matrix ${\cal M}_*$ is well approximated by
\ba
{\cal M}_* \approx \frac{1}{g} 
\left(
\begin{array}{cc}
|A|^2 + |B|^2, & 2AB \\
2 \bar{A}\bar{B}, & |A|^2 + |B|^2 \\
\end{array}
\right)_*. 
\ea
The two eigenvalues are 
\ba
\frac{1}{g} (|A| \pm |B|)_*^2 = \frac{1}{g} \left( |W''- (\partial \partial_F V_{{\rm 1-loop}})| \pm 
|(\partial \partial_{\bar{F}}V_{{\rm 1-loop}})| \right)_*^2, 
\ea
respectively, ensuring the positivity of (\ref{massmatrix}). 

\begin{figure}[htbp]
 \begin{center}
  \includegraphics[width=80mm]{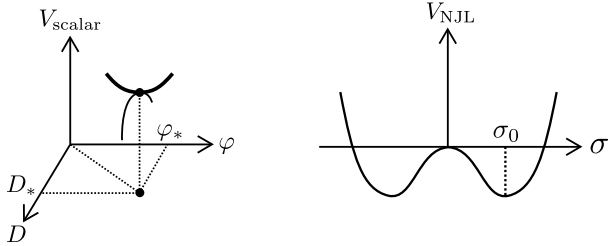}
 \end{center}
 \caption{Comparison of $V_{{\rm scalar}}$ around the stationary value $(D_*, \varphi_*)$ with $V_{{\rm NJL}}$.} 
 \label{potential}
\end{figure}

For $V_0$, $y_L =D, \vec{y}_R=(\varphi, \bar{\varphi})$, 
\ba
&&M_{RR_*} = 
\left(
\begin{array}{cc}
\partial^2 V_0, & \partial \bar{\partial} V_0 \\
\bar{\partial} \partial V_0, & \bar{\partial}^2 V_0 \\
\end{array}
\right)_*, \qquad 
M_{RL_*} = 
\left(
\begin{array}{c}
\partial \partial_D V_0 \\
\bar{\partial} \partial_D V_0 \\
\end{array}
\right)_*, \\
&&M_{LR_*} =M_{RL}^*, \qquad M_{LL_*} = \partial_D^2 V_{0*}.  
\ea
We know that the $D$ profile of $V_0(D, \varphi, \bar{\varphi})$ near the stationary point is 
convex to the top and 
we fit this by
\ba
V_0 = V_h(\varphi, \bar{\varphi}) - \frac{\alpha(\varphi, \bar{\varphi})}{2} \left( D - D_*(\varphi, \bar{\varphi}) \right)^2 
+ {\cal O}((D - D_*(\varphi, \bar{\varphi}))^4). 
\ea
Here $\alpha$ is a positive real function of $\varphi, \bar{\varphi}$ and $V_h(\varphi, \bar{\varphi}) = V_0(D_*(\varphi, \bar{\varphi}), \varphi, \bar{\varphi})$. 
One can check
\ba
-M_{RL_*} M_{LL_*}^{-1} M_{LR*} = \alpha_* 
\left(
\begin{array}{c}
\partial D_*\\
\bar{\partial} D_*\\
\end{array}
\right)_*
((\partial D_*), (\bar{\partial} D_*))_*,
\ea
while
\ba
M_{RR*} = 
\left(
\begin{array}{cc}
\partial^2 V_h & \partial \bar{\partial} V_h \\
\bar{\partial} \partial V_h & \bar{\partial}^2 V_h \\
\end{array}
\right)_*
-\alpha 
\left(
\begin{array}{cc}
(\partial D_*)^2 & |\partial D_*|^2 \\
|\partial D_*|^2 & (\bar{\partial} D_*)^2 \\
\end{array}
\right)_*
\ea
and
\ba
\delta^2 V_{0*} &=& \frac{1}{2} \delta \vec{y}_R 
\left( M_{RR*} - M_{RL*} M_{LL*}^{-1} M_{LR*} \right)_*
\delta \vec{y}_R \nonumber \\
&=& \delta \vec{y}_R^\dag 
\left(
\begin{array}{cc}
\partial \bar{\partial} V_h & \partial^2 V_h \\
\bar{\partial}^2 V_h & \partial \bar{\partial} V_h \\
\end{array}
\right)_*
\delta \vec{y}_R 
\equiv \delta \vec{y}_R^\dag {\cal M}_{h*} \vec{y}_R. 
\label{V0massmatrix}
\ea
The entire contribution of the second variation  
  $\delta^2 V_* = \delta^2 {\cal V}_* + \delta^2 V_{0*}$ to the leading order in the Hartree-Fock approximation 
  is given by Eqs.(\ref{massmatrix}), (\ref{V0massmatrix}). 
The mass of the scalar gluons squared is obtained by multiplying the combined mass matrix by $g^{-1}_*$:
\ba
  g^{-1}_* ({\cal M}_* + {\cal M}_{h*}), 
\ea
generalizing the tree formula.  
  In practice, we just need a well-approximated formula  valid in the region we work with and one can invoke
  the $U(1)$ invariance to ensure that the two eigenvalues of the complex scalar gluons are degenerate. 
  Let us, therefore, use the expression
\ba
&&\frac{1}{g} |W''- (\partial \partial_F V_{{\rm 1-loop}})|_*^2  + \partial \bar{\partial} V_{h*} \nonumber \\
&=& \left|
\frac{\varphi_*}{M} 
-iN^2 \left( \frac{\varphi_*}{M} \right)^2 
\left[
\frac{A(\varepsilon, \gamma)}{32\pi^2} 
\left\{
(\lambda_{0*}^{+})^4 + (\lambda_{0*}^{-})^4 + \frac{2}{\varphi_*/M} -2 -\frac{3}{4} \Delta_{0*}^2 +\frac{1}{8} \Delta_{0*}^4
\right\} \right. \right. \nonumber \\
&& \left. \left. -\frac{1}{32\pi^2} 
\left\{
(\lambda_{0*}^{+})^4 \log (\lambda_{0*}^{+})^2 + (\lambda_{0*}^{-})^4 \log (\lambda_{0*}^{-})^2 
\right. \right. \right. \nonumber \\
&& \left. \left. \left. + \frac{3(1+\frac{\Delta_{0*}^2}{2})}{\sqrt{1+\Delta_{0*}^2}} 
\left((\lambda_{0*}^{-})^3 \left( \log (\lambda_{0*}^{-})^2 + \frac{1}{2} \right) 
- (\lambda_{0*}^{+})^3 \left( \log (\lambda_{0*}^{+})^2 + \frac{1}{2} \right) \right)
\right\} 
-\frac{2}{32\pi^2} \frac{1}{\varphi_*/M}
\right] \right|^2 M^2 \nonumber \\
&& + 2N^2 \left(\frac{\varphi_*}{M} \right)^2 
\left[
2\left( c' +\frac{1}{64\pi^2} \right) \Delta_{0*}^2 
+ \frac{2}{32\pi^2} \left( \frac{\tilde{A}}{8} -(\lambda_{0*}^+)^4 \log (\lambda_{0*}^+) - (\lambda_{0*}^-)^4 \log(\lambda_{0*}^-)^2 \right) 
\right. \nonumber \\
&& \left. 
+ \frac{{\rm Im}(i+\Lambda)}{N^2} \frac{1}{\varphi_*/M}
\right] M^2
\label{stabcriterion}
\ea
to check the local stability of the potential and the mass. 
The above expression is obtained for our simple example of ${\cal F}$ and $W$
\ba
{\cal F} = \frac{i}{2} \varphi^2 + \frac{1}{3!M} \varphi^3, \qquad
W = m^2 \varphi + \frac{1}{3!} \varphi^3,
\ea 
and the real case $\Delta_0=\bar{\Delta}_0$ is applied. 
Using the numerical analyses carried out in the last subsection, 
we have made a list of data on Eq. (\ref{stabcriterion}). 
\begin{table}[htb]
\begin{center}
  \begin{tabular}{|c|c|c|c|l|} 
  \hline
$c'+\frac{1}{64\pi^2}$ & $\tilde{A}/(4\cdot32\pi^2)$ & $\Delta_{0*}$  
& $\varphi_*/M~(-\frac{N^2}{{\rm Im}(i+\Lambda)})$ & scalar gluon mass \\ 
\hline
0.002 & 0.0001 & 0.477 & 0.707~(10000) & 0.4998 + 0.0056 $N^2$ + $8.607 \times 10^{-7} N^4$ \\
\hline
0.003 & 0.001 & 1.3623 & 0.8639~(2000) & 0.7463 + 0.0106 $N^2$ + $2.653 \times 10^{-4} N^4$  \\
0.003 & 0.001 & 1.3623 & 0.5464~(5000) & 0.2986 + 0.0008 $N^2$ + $4.694 \times 10^{-5} N^4$ \\
0.003 & 0.001 & 1.3623 & 0.3863~(10000) & 0.1492 $-$ 0.0024 $N^2$ +$7.235 \times 10^{-5} N^4$ \\
 \hline
  \end{tabular}
\end{center}
\caption{Samples of numerical values for the scalar gluon masses.}
\label{sgluonmass}
\end{table}
Except for the last case in the Table \ref{sgluonmass}, 
the scalar gluon masses squared are found to be positive for any $N$, 
which implies that our stationary points are locally stable. 
Even in the last case, the stability is ensured for small $N$. 
In these data, we have checked that 
the inequalities $|(\partial_F \partial_{\bar{F}} V_{{\rm 1-loop}})|_*, |(\partial_F^2 V_{{\rm 1-loop}})|_* \ll g_*$ are in fact  satisfied. 
As a summary of our understanding, 
a schematic figure is drawn in Fig. \ref{potential}, which illustrates the local stability of the scalar potential 
at the vacuum of dynamically broken supersymmetry in comparison with the well-known NJL potential.

\subsection{Choice of regularization and subtraction scheme} 
In this paper, we have considered the theory specified by the general  ${\cal N}=1$ supersymmetric lagrangian Eq. (\ref{KtauW}), 
have regularized the theory by the supersymmetric dimensional regularization (dimensional reduction) 
and have subtracted the part of the $1/\epsilon$ poles of the regularized one-loop effective action in Eq. (\ref{1loopeffpot}) 
by the supersymmetric subtraction scheme defined by the condition  Eq. (\ref{rencond}).
The original infinity is transmuted into the infinite constant $\Lambda$ 
which is the coefficient of the counterterm and the effective potential has been recast to describe 
the behavior of the theory well below the UV cutoff residing in the prepotential function. 

We now make brief comments on other regularizations and subtraction schemes which we did not employ in this paper. 
The relativistic momentum cutoff is a natural choice of the NJL theory as we mentioned earlier 
but regularizing the integral Eq. (\ref{1loopeffaction}) by the momentum cutoff leads us to a rather unwieldy expression. 
See Ref. \cite{imaru2}. 
Unlike supersymmetric dimensional reduction \cite{Siegel},  
the momentum cutoff ${\it per se}$, while preserving the equality between the Bose and Fermi degrees of freedom, 
does not have a firm basis on the regularized action which the supersymmetry algebra acts on. 
Moreover, as is clear from $(A.1)$ of Ref. \cite{imaru2}, 
the result violates the positivity of the effective potential in the vicinity of the origin in the $\Delta$ profile. 
This violation is a necessity in the broken chiral symmetry of the NJL theory 
but here it contradicts with the positive semi-definiteness of energy that the rigid supersymmetric theory possesses. 
Turning to the choice of the subtraction scheme, 
one might also like to apply the ``(modified) minimal subtraction scheme" in our one-loop integral Eq. (\ref{1loopeffpot}). 
While we do not know how to justify this prescription here, the subsequent analyses proceed almost in the same way 
and the main features of the equations obtained from our variational analyses and the conclusions are unchanged. 

\section{Lifetime of metastable SUSY breaking vacuum}

Combining the two facts that the trivial solution $\Delta_0=0$ of the gap equation is also a trivial solution 
and the energy in rigid SUSY theories is positive semi-definite leads us that our SUSY breaking vacuum is a local minimum. 
For our mechanism to be viable, 
we have to show that our SUSY breaking vacuum is sufficiently long-lived during the decay into the true vacuum with $\Delta_0=0$; 
in other words, the lifetime of our vacuum must be much longer than the age of universe. 
Taking into account the nonvanishing F-term VEV induced by D-term VEV as well, 
we carry out an order estimate of the lifetime of our SUSY breaking vacuum. 
Neglecting ${\cal O}(1)$ quantities, we have
\ba
\frac{1}{2} \langle D^0 \rangle^2 \sim {\cal O}(m_s^2 \Lambda^2), 
\quad \langle V_{{\rm 1-loop}} \rangle \sim {\cal O} \left(\frac{\alpha}{4\pi} N^2 m_s^4 \right)  
\sim {\cal O} \left(m_s^4 \right) 
\label{VEV1}
\ea
where $\Lambda$ is a cutoff scale. 
Plugging these VEVs into Eq.~(4.1) in \cite{imaru2} 
\ba
|\langle F^0 \rangle|^2 + \frac{m_s}{\langle g^{00} \partial_0 g_{00} \rangle} \langle F^0 \rangle 
+ \frac{1}{2} \langle D^0 \rangle^2 
+ 2 \langle g^{00} \rangle \langle V_{{\rm 1-loop}} \rangle =0. 
\ea
leads to 
\ba
\langle F^0 \rangle \sim {\cal O}(m_s \Lambda)
\ea
provided $m_s \ll \Lambda$. 

The decay rate of our vacuum to the true one is controlled by the factor $\exp[ - |\langle \Delta \phi \rangle|^4/\langle \Delta V \rangle]$ as seen in Ref. \cite{ISS},  
where $\langle \Delta \phi \rangle, \langle \Delta V \rangle$ are the scalar field distance and the potential height between two vacua. 
These two quantities are estimated as follows. 
\ba
&&\langle F^0 \rangle 
= - \langle g^{00}\rangle \langle \overline{\partial_0 \partial_0 W} \rangle  \langle \Delta \overline{\phi}^0 \rangle 
=- m_s \langle \Delta \overline{\phi}^0 \rangle \Rightarrow \langle \Delta \overline{\phi}^0 \rangle \sim {\cal O}(\Lambda), \\
&&\langle \Delta V \rangle = |\langle F^0 \rangle|^2 + \frac{1}{2} \langle D^0 \rangle^2 + \langle V_{{\rm 1-loop}} \rangle \sim {\cal O}(m_s^2 \Lambda^2). 
\ea
Using these results, the requirement of the longevity of our metastable vacuum is given by the condition  
\ba
\frac{| \langle \Delta \phi^0 \rangle|^4}{\langle \Delta V \rangle} \sim {\cal O}\left( \frac{\Lambda^2}{m_s^2} \right) \gg 1,
\ea
which is always satisfied as long as $m_s \ll \Lambda$. 

\section{Higgs Mass}
In order to realize the observed Higgs mass 126 GeV in the MSSM, 
SUSY breaking scale would be higher since the Higgs boson requires large radiative corrections from the top squarks.  
Also, we have no signals for SUSY particles from the experiments at Large Hadron Collider (LHC). 
Thus, the naturalness of MSSM becomes worse and worse.  
It is well known fact that Higgs mass in the MSSM at tree level is smaller than the Z-boson mass, 
but it can be avoided if we consider extensions of the MSSM. 

In this section, we investigate implications of the mechanism of DDSB uncovered in \cite{imaru1, imaru2, imaru3}, 
 coupling the system to the MSSM Higgs sector which includes the $\mu$ and $B\mu$ terms \cite{imaru4}.
The pair of Higgs doublet superfields $H_u, H_d$ is taken to be charged under the overall $U(1)$:
\ba
{\cal L}_{{\rm Higgs}} &=& \int d^4 \theta 
 \left[
 H_u^\dag e^{-g_Y V_1 - g_2 V_2 - 2e_u V_0} H_u 
 +H_d^\dag e^{g_Y V_1 - g_2 V_2 - 2e_d V_0} H_d 
 \right]  \nonumber \\
&& + \left[ \left(  \int d^2 \theta \mu H_u \cdot H_d \right)  
- B\mu H_u \cdot H_d + {\rm h.c.} \right] \;.
\label{HiggsLagrangian}
\ea
We have adopted notation $X \cdot Y \equiv \epsilon_{AB} X^A Y^B = X^A Y_A = - Y \cdot X$,
$\epsilon_{12} = - \epsilon_{21} = \epsilon^{21} = - \epsilon^{12} =1$. 
$V_{1,2,0}$ are vector superfields of the SM gauge group and that of the overall $U(1)$ respectively 
 and the corresponding gauge couplings are denoted by $g_{Y,2}$ and $e_{u,d}$ respectively.  
Unlike the MSSM case, the soft scalar Higgs masses $m_{H_u}^2 |H_u|^2, m_{H_d}^2 |H_d|^2$ are not introduced 
 since they are induced by $D$-term contributions in our framework.  

To simplify the analysis in what follows while keeping the essence, 
 we adopt the simplest prepotential and superpotential exploited in Eqs. (\ref{F}) and (\ref{W}) 
 of $5 \times 5$ complex matrix scalar superfield  $\varphi$ :
 \ba
{\cal F} =\frac{c}{2N}  {\rm tr} \varphi^2 + \frac{1}{3!MN} {\rm tr} \varphi^3\;, \;\;
W = \frac{m^2}{N} {\rm tr} \varphi + \frac{d}{3!N} {\rm tr} \varphi^3\;,
\label{f&W}
\ea
where $c$ is a pure imaginary number as discussed above, and $m,M$ are mass parameters. 
Here $N=5$ and $M$ (real number) sets the scale in the prepotential, which is the cutoff scale.
    
We embed the generators of the gauge group into the bases which expand
  $\varphi$:
  \ba
 \varphi \equiv
\left(
\begin{array}{cc}
T_8 & 0 \\
0 &  T_3\\
\end{array}
\right)
   +  \sqrt{\frac{3}{5}} Y 
\left(
\begin{array}{cc}
 -\frac{1}{3} {\bf 1}_3 & 0 \\
0 &   \frac{1}{2} {\bf 1}_2   \\
\end{array}
\right)
  + \frac{ {\bf 1}_5}{\sqrt{10}} S \;, \;\;
T_3 = \sum_{a=1}^{3} T^a \left( \frac{\sigma^a}{2} \right)\;\;.
\label{varphiexpansion}
\ea
  We have represented  the overall $U(1)$ and $U(1)_Y$  generators to be proportional to the unit matrix 
    and  the traceless diagonal generator respectively. 
  We analyze the case in which  only $S$  receives its VEV,
     namely, the unbroken $U(5)$ vacuum of the superpotential. 
We will make a comment for those cases in which these do not hold,
   which lead to the kinetic mixing.
We drop octet $T_8$ as it is irrelevant to the analysis below. 
    
After a simple calculation, we obtain  the non-vanishing prepotential derivatives
 \ba
 {\cal F}_{aa} &=&  \frac{c}{10} + \frac{3}{3! 5 \sqrt{10} M} \left( \sqrt{\frac{3}{2}} Y + S \right), \;
  {\cal F}_{00} = \frac{c}{10} + \frac{3 S}{3! 5 \sqrt{10} M}, \;  \nonumber \\
    {\cal F}_{YY} &=& \frac{c}{10} + \frac{3}{3! 5 \sqrt{10} M} \left( \sqrt{\frac{1}{6}} Y + S \right), \; 
    {\cal F}_{a0} =  \frac{3}{3! 5 \sqrt{10} M}  T^a, \;  \nonumber \\
     {\cal F}_{aY} &=&  \frac{3}{3! 5 \sqrt{10} M} \sqrt{\frac{3}{2}}
      T^a, \;   {\cal F}_{0Y} =  \frac{3}{3! 5 \sqrt{10} M} Y\;,
 \label{prep2nd}
 \ea
 their VEV's
 \ba
 \langle {\cal F}_{aa} \rangle  &=&  \langle {\cal F}_{YY} \rangle = \langle {\cal F}_{00} \rangle  =
  \frac{c}{10} + \frac{3}{3! 5 \sqrt{10} M}  \langle S \rangle , \; \nonumber \\
  \langle {\cal F}_{a0} \rangle  &=&  \langle {\cal F}_{aY} \rangle = \langle {\cal F}_{0Y} \rangle =0\;,
 \label{prep2ndvev}
 \ea
 and the derivatives of the superpotential
\ba
 \partial_a W &=&  \frac{3 d}{3! 5 \sqrt{10}} T^a \left( \sqrt{\frac{3}{2}} Y + S \right), \; \nonumber \\
 \partial_0 W &=&   \frac{m^2}{\sqrt{10}} +
  \frac{3 d}{3! 10 \sqrt{10}}  \left(  \sum_{a} T^aT^a +  Y^2 + S^2 \right), \; \nonumber \\
 \partial_Y W &=&  \frac{3 d}{3! 5 \sqrt{10}}  \left( \frac{3}{4} \sum_{a} T^aT^a +
 \frac{1}{4} Y^2 + \sqrt{\frac{3}{2}} S Y \right)\;.
\label{derisup}
\ea
 We choose $c= 10i$ but $\langle S \rangle $ is complex, not necessarily real.  

In this paper, we add Eq. (\ref{HiggsLagrangian}) to Eq. (\ref{KtauW})
 and consider a part 
 relevant to 126 GeV Higgs
\ba
{\cal L} &=&  {\cal L}_{{\rm Higgs}} 
 + \int d^2 \theta
           {\rm Im} \frac{1}{2} 
           {\cal F}_{ab}(\Phi^a)
           {\cal W}^{\alpha a} {\cal W}^b_{\alpha}  \nonumber \\
&=& {\cal L}_{{\rm Higgs}} 
+ \frac{1}{4} 
\left[ 
\int d^2 \theta ( {\cal W}^a {\cal W}^a + {\cal W}^Y {\cal W}^Y + {\cal W}^0 {\cal W}^0 ) 
 + {\rm h.c.} \right] \nonumber \\
&&+\frac{1}{4} \left[ 
\int d^2\theta (
{\cal F}_{aaY} Y {\cal W}^a {\cal W}^a 
+ {\cal F}_{aa0} S {\cal W}^a {\cal W}^a 
+ {\cal F}_{YYY} Y {\cal W}^Y {\cal W}^Y 
+ {\cal F}_{YY0} S {\cal W}^Y {\cal W}^Y 
\right. \nonumber \\
&& \left. 
+ {\cal F}_{000} S {\cal W}^0 {\cal W}^0 
+ {\cal F}_{a0a} T^a {\cal W}^a {\cal W}^0 
+ {\cal F}_{aYa} T^a {\cal W}^a {\cal W}^Y 
+ {\cal F}_{0YY} Y {\cal W}^0 {\cal W}^Y 
)
+{\rm h.c.}
\right]. 
\label{ConcreteL}
\ea
The third prepotential derivatives, which are now real numbers, 
can be read off from Eq. (\ref{prep2ndvev}). 

In our analysis, we take that the value of $D^0$ VEV is determined essentially 
 by our Hartree-Fock approximation in \cite{imaru3}. 
This source of supersymmetry breaking is then fed to the Higgs sector 
 and its effects are given by a tree level analysis. 
We will argue the validity of this procedure below.    

\subsection{Higgs potential and variations}
Let us extract the part relevant to the Higgs potential in (\ref{ConcreteL}). 
\ba
{\cal L}_{{\rm pot}} &=& 
|F_{H_u}|^2 + \left(- \frac{g_Y}{2} D^Y -e_u D^0 \right) |H_u|^2 -g_2 H_u^\dag \sum_a D^a \frac{\sigma^a}{2} H_u 
\nonumber \\
&& 
+ |F_{H_d}|^2 + \left( \frac{g_Y}{2} D^Y -e_d D^0 \right) |H_d|^2 -g_2 H_d^\dag \sum_a D^a \frac{\sigma^a}{2} H_d
\nonumber \\
&& 
- \left( \mu H_u \cdot F_{H_u} + \mu F_{H_d} \cdot H_d +B\mu H_u \cdot H_d + {\rm h.c.} \right) 
+ \frac{1}{2} \left( \sum_a D^a D^a + (D^Y)^2 + (D^0)^2 \right) 
\nonumber \\
&& 
+ \frac{1}{2} \sum_{A,B,C=a,Y,0} {\rm Im} ({\cal F}_{ABC} \varphi^C) D^A D^B 
+ \Gamma^{{\rm 1-loop}}(D^0)
\label{Dpot}
\ea
where $\varphi^C=(T^a, Y, S)$. 
The one-loop part of the effective potential in \cite{imaru1, imaru3} is denoted by $\Gamma^{{\rm 1-loop}}(D^0)$. 
Fermionic backgrounds are not needed in the potential analysis of Higgs and are not included in Eq. (\ref{Dpot}). 

Let us vary ${\cal L}_{{\rm pot}}$
with respect to the auxiliary fields, 
 replacing $\varphi^C$ by their VEV $\langle \varphi^C \rangle =(0, 0, \langle S \rangle)$. 
\ba
\delta D^a: && 0 = (1+{\rm Im}{\cal F}''' \langle S \rangle)D^a 
-g_2 H_u^\dag \frac{\sigma^a}{2} H_u -g_2 H_d^\dag \frac{\sigma^a}{2} H_d, 
\label{Da} \\
\delta D^Y: && 0 = (1+{\rm Im}{\cal F}''' \langle S \rangle)D^Y 
- \frac{g_Y}{2} |H_u|^2  + \frac{g_Y}{2} |H_d|^2, 
\label{DY} \\
\delta D^0: && 0 = (1+{\rm Im}{\cal F}''' \langle S \rangle)D^0 
-e_u |H_u|^2 -e_d |H_d|^2 + \frac{\partial \Gamma^{{\rm 1-loop}}(D^0)}{\partial D^0}. 
\label{D0} 
\ea
Note that ${\cal F}_{aa0}={\cal F}_{YY0}={\cal F}_{000} \equiv {\cal F}'''$  and 
 that Eq. (\ref{D0}) with $e_u = e_d = 0$ is in fact the gap equation of \cite{imaru1, imaru3}. 
 Eliminating the auxiliary fields (approximately), we obtain Higgs potential
\ba
V_{{\rm Higgs}} &=& \frac{g_2^2}{2(1+{\rm Im}{\cal F}'''\langle S \rangle)} 
\left( H_u^\dag \frac{\sigma^a}{2} H_u + H_d^\dag \frac{\sigma^a}{2} H_d \right)^2 
+ \frac{g_Y^2}{8(1+{\rm Im}{\cal F}'''\langle S \rangle)} \left( |H_u|^2 - |H_d|^2 \right)^2 
\nonumber \\
&&
+ \frac{1}{2(1+{\rm Im}{\cal F}'''\langle S \rangle)} 
\left( e_u |H_u|^2 + e_d |H_d|^2 - \left. \frac{\partial \Gamma^{{\rm 1-loop}}(D^0)}{\partial D^0} \right|_{D^0=D^{0*}} \right)^2 
\nonumber \\
&&
+ |\mu|^2 (|H_u|^2 + |H_d|^2) + (B \mu H_u \cdot H_d + {\rm h.c.}). 
\ea 
Here we have denoted by $D^{0*}$ the solution to Eq.(\ref{D0}) the improved gap equation. 
The deviation $\delta D^{0*}$ of the value from $D^{0*}$ in \cite{imaru3} is in fact small 
by the ratio of electroweak scale and SUSY breaking scale. 
Therefore, we approximate the solution to the improved gap equation 
 by the value of $D^{0*}$ in \cite{imaru3} denoted as $\langle D^0 \rangle$. 
Taking into account the fact that ${\rm Im}{\cal F}'''\langle S \rangle \sim \langle S \rangle/M \ll 1$, 
 we neglect the term ${\rm Im}{\cal F}'''\langle S \rangle$ at the leading order. 
The resulting Higgs potential at the leading order is given by
\ba
V_{{\rm Higgs}} &\simeq& 
\frac{g_2^2}{2} \left( H_u^\dag \frac{\sigma^a}{2} H_u + H_d^\dag \frac{\sigma^a}{2} H_d \right)^2 
+ \frac{g_Y^2}{8} \left( |H_u|^2 - |H_d|^2 \right)^2 
\nonumber \\
&&
+ \frac{1}{2} \left( e_u |H_u|^2 + e_d |H_d|^2 - \langle D^0 \rangle \right)^2 
+ |\mu|^2 (|H_u|^2 + |H_d|^2) + (B \mu H_u \cdot H_d + {\rm h.c.})
\nonumber \\
&=& 
\frac{g_2^2 + g_Y^2}{8} 
\left[
|H_u^0|^2 - |H_d^0|^2
\right]^2
+ \frac{1}{2} 
\left(
e_u |H_u^0|^2 + e_d |H_d^0|^2 - \langle D^0 \rangle 
\right)^2 \nonumber \\
&&+ |\mu|^2 \left( |H_u^0|^2 + |H_d^0|^2\right) 
- \left( B \mu H_u^0 H_d^0 + {\rm h.c.} \right) 
\nonumber \\
&=&
\frac{g_2^2 + g_Y^2}{32} v^4 c_{2\beta}^2 
+ \frac{v^2}{2} 
\left[ 
\mu^2 - B \mu s_{2\beta}
\right]  
+ \frac{1}{8}  
\left(
( e_u s_\beta^2 + e_d c_\beta^2)v^2 - 2 \langle D^0 \rangle  
\right)^2 
\ea 
where we have restricted the potential to the CP-even neutral sector 
of Higgs doublets $H_u=(H_u^+, H_u^0)^T$, $H_d=(H_d^0, H_d^-)^T$ 
in the second line 
 since we are interested in the Higgs mass. 
In the last line,   
 the neutral components of Higgs fields are parametrized as
\ba
&&H_u^0 = \frac{1}{\sqrt{2}} \left[ s_\beta (v+h) + c_\beta H + i (c_\beta A - s_\beta G^0 ) \right], \\
&&H_d^0 = \frac{1}{\sqrt{2}} \left[ c_\beta (v+h) - s_\beta H + i (s_\beta A + c_\beta G^0 ) \right]
\ea
where we use the shorthand notations: 
\ba
s_\beta \equiv \sin \beta, \quad c_\beta \equiv \cos \beta, \quad t_\beta \equiv \tan \beta, \quad 
s_{2\beta} \equiv \sin 2\beta, \quad c_{2\beta} \equiv \cos 2 \beta. 
\ea
$G^0$ is the would-be Nambu-Goldstone boson eaten as the longitudinal component of $Z$-boson. 
The VEV of Higgs field is $v \simeq 246$~GeV and $\frac{g_Y^2+g_2^2}{4}v^2=M_Z^2$ in this convention.

\subsection{Estimate of the Higgs mass}
We are now ready to calculate Higgs mass. 
As in the MSSM, 
 the minimization of the scalar potential $\partial V_{{\rm Higgs}}/\partial v^2 = \partial V_{{\rm Higgs}}/\partial \beta = 0$ 
 allows us to express $\mu$ and $B\mu$ in terms of other parameters. 
\ba
\mu^2 + \frac{M_Z^2}{2} &=& 
\frac{1}{2c_{2\beta}} \left( (e_u s_\beta^2 + e_d c_\beta^2 )v^2 - 2 \langle D^0 \rangle \right) 
\left(
e_u s_\beta^2 - e_d c_\beta^2
\right), \\
M_A^2 \equiv \frac{2B \mu}{s_{2\beta}} 
&=& 2 \mu^2 
+ \frac{e_u+e_d}{2} \left( (e_u s_\beta^2 + e_d c_\beta^2 ) v^2 - 2 \langle D^0 \rangle \right) \nonumber \\
&=& -M_Z^2 + \frac{e_u-e_d}{2 c_{2\beta}} 
\left( (e_u s_\beta^2 + e_d c_\beta^2 ) v^2 - 2 \langle D^0 \rangle \right). 
\ea
It is straightforward to obtain the mass matrix for CP-even Higgs from the second derivative of the potential,  
\ba
{\cal M}^2 =
\left(
\begin{array}{cc}
m^2_{hh} & m^2_{hH} \\
m^2_{hH} & m^2_{HH} \\
\end{array}
\right)
\ea
where each component is given by
\ba
m_{hh}^2 &=& M_Z^2 c_{2\beta}^2 + v^2 \left( e_u s_\beta^2 + e_d c_\beta^2 \right)^2, \\
m^2_{HH} &=& M_A^2 + M_Z^2 s_{2\beta}^2 
+ v^2 s_{2\beta}^2 
\left(
\frac{e_u-e_d}{2} 
\right)^2,\\
m_{hH}^2 &=& -M_Z^2 s_{2\beta} c_{2\beta} 
+ v^2 s_{2\beta} 
\left( e_u s_\beta^2 + e_d c_\beta^2 \right)
\left( \frac{e_u-e_d}{2} \right).  
\ea
The eigenvalues of this mass matrix are found as
\ba
\frac{1}{2} \left[ m_{hh}^2 + m^2_{HH} \pm \sqrt{ (m_{hh}^2 - m^2_{HH})^2 + 4m_{hH}^4 } \right] 
\ea
and the lightest CP-even Higgs mass is 
\ba
m_{{\rm Higgs}}^2 &=& \frac{1}{2} \left[ m_{hh}^2 + m^2_{HH} - \sqrt{ (m_{hh}^2 - m^2_{HH})^2 + 4m_{hH}^4 } \right]. 
\ea 
In order for the $\mu$-term to be allowed in the superpotential, 
 we must have a condition $e_u+e_d=0$ which is also required from an anomaly cancellation condition for the overall $U(1)$. 
Then, the Higgs mass can be expressed as
\ba
m_{{\rm Higgs}}^2 &=& 
\frac{1}{2} \left[ 
M_Z^2 + M_A^2 + e_u^2 v^2  
-\sqrt{\left( M_A^2-M_Z^2 c_{4\beta} - c_{4\beta} e_u^2 v^2 \right)^2 
+ s_{4\beta}^2 \left( M_Z^2 + e_u^2 v^2 \right)^2}
\right] \nonumber \\
&=& 
\frac{1}{2} \left[ 
\tilde{M}_Z^2 + M_A^2 
-\sqrt{\left( 
\tilde{M}_Z^2 + M_A^2 
 \right)^2 
 -4 \tilde{M}_Z^2 M_A^2c_{2\beta}^2 
 }
\right] 
\label{Hmass}
\ea
where 
$\tilde{M}^2_Z \equiv M_Z^2  + e_u^2 v^2$. 
It is interesting to see the correspondence between our expression of Higgs mass (\ref{Hmass}) and that in the MSSM,
\ba
m_{{\rm MSSM~Higgs}}^2 &=& 
\frac{1}{2} \left[ 
M_Z^2 + M_A^2 
-\sqrt{\left( M_Z^2 + M_A^2 \right)^2 
 -4 M_Z^2 M_A^2c_{2\beta}^2 }
\right]. 
\ea 
As in the case of MSSM, the upper bound of Higgs mass can be obtained 
 by taking a decoupling limit $M_A^2 \to \infty$, 
\ba
m^2_{{\rm Higgs}} \to \tilde{M}_Z^2 c_{2\beta}^2. 
\ea
$\tilde{M}_Z$ can be large enough by taking ${\cal O}(1)$ charge $e_u$ 
\ba
\tilde{M}_Z \sim \sqrt{(90~{\rm GeV})^2 + \left(246~{\rm GeV} \right)^2} \sim 262~{\rm GeV}. 
\ea
Let us go back to the minimization conditions of Higgs potential with $e_u+e_d=0$,
\ba
\mu^2+\frac{M_Z^2}{2} &=& \frac{e_u}{2c_{2\beta}} \left( -c_{2\beta}e_u v^2 - 2 \langle D^0 \rangle \right), \\
M_A^2 &=& 2\mu^2 
 = -M_Z^2 - \frac{e_u}{c_{2\beta}} 
\left( e_u c_{2\beta}  v^2 - 2 \langle D^0 \rangle \right)
\ea
which leads to
\ba
M_Z^2 + M_A^2 = -\frac{e_u}{c_{2\beta}} \left( c_{2\beta} e_u v^2 + 2 \langle D^0 \rangle \right).
\label{minimization}
\ea
In order to satisfy this condition, 
 the dominant part in the right-hand side of (\ref{minimization}) 
 $e_u \langle D^0 \rangle/c_{2\beta}$ is required to be negative. 

Using these conditions, we can eliminate $M_A^2$ in Higgs mass (\ref{Hmass}). 
\ba
m^2_{{\rm Higgs}} &=& \frac{1}{2} 
\left[
-\frac{2e_u}{c_{2\beta}} \langle D^0 \rangle 
- \sqrt{\left(-\frac{2e_u}{c_{2\beta}} \langle D^0 \rangle \right)^2 
+ 8 c_{2\beta} e_u \tilde{M}_Z^2 \langle D^0 \rangle + 4c_{2\beta}^2 \tilde{M}^4_Z}
\right] 
\nonumber \\
&\simeq&
\tilde{M}^2_Z c_{2\beta}^2 \left( 1+\frac{c_{2\beta} \tilde{M}^2_Z}{2e_u \langle D^0 \rangle} s_{2\beta}^2 \right) 
\ea
where the approximation $\langle D^0 \rangle \gg \tilde{M}_Z^2$ is applied in the second line. 

A plot for 126 GeV Higgs mass as a function of $\cos 2\beta$ and $e_u$ is shown below. 
Here we have taken $e_u < 0$ and $\cos 2\beta>0$ to satisfy the condition $e_u \langle D^0 \rangle/c_{2\beta}<0$.
We can immediately see that 126 GeV Higgs mass is realized by ${\cal O}(1)$ charge $e_u$, 
 namely without fine-tuning of parameters. 
Also, we found that the result is insensitive to the values of $D$-term VEV. 
This fact is naturally expected from the non-decoupling nature of Higgs mass.

\begin{figure}[htbp]
 \begin{center}
  \includegraphics[width=70mm]{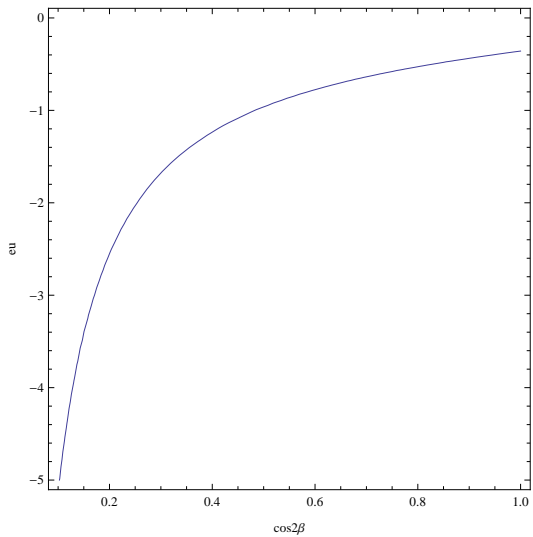}
 \end{center}
 \caption{A plot for 126 GeV Higgs mass as a function of $\cos 2\beta$ and $e_u$. 
 The result is insensitive to the values of D-term VEV.}
 \label{126GeV}
\end{figure}

\section{Summary}
In this paper, we reviewed our recently proposed mechanism of 
$D$-term triggered dynamical SUSY breaking \cite{imaru1, imaru2, imaru3, imaru4}. 
The nonvanishing D-term VEV is dynamically realized 
as a nontrivial solution of the gap equation in the Hartree-Fock approximation 
as in the case of NJL model and BCS superconductivity. 
In our mechanism, since the gauge sector is extended to be ${\cal N} = 2$ supersymmetric, 
gaugino becomes massive by the D-term VEV through the Dirac mass term with ${\cal N}=2$ partner fermion of gaugino. 
Our mechanism can be directly applied to Dirac gaugino scenario which much attention has been paid to. 

A systematic analysis of the scalar potential was performed 
by treating the order parameters of SUSY breaking $D$ and $F$, 
and the adjoint scalar field as the background fields. 
It was shown numerically that SUSY is indeed broken dynamically 
and our meta-stable vacuum is locally stable. 
The lifetime of our meta-stable vacuum was also shown to be sufficiently long-lived. 

As a phenomenological application,  
we have discussed how an observed Higgs mass can be realized in the context of DDSB 
and have shown that it is naturally realized by an additional overall $U(1)$ D-term contribution to Higgs mass.

\section*{Acknowledgment}
The author would like to thank H. Itoyama for the series of collaborations.

\end{document}